\documentclass[12pt]{article}
\usepackage{epsf}

\def\ltap{\ \raisebox{-.4ex}{\rlap{$\sim$}} \raisebox{.4ex}{$<$}\ }

\begin{document}

%%%%%%%%%%%%%%%%%%%%%%%%%%%%%%%%%%%%%%%%
%%%%%%% REMOVE THIS PART FOR PROCEEDINGS %%%%%%%%%%%%%%%%%%%%%%%%%%%%%%%%

\begin{titlepage}

\begin{flushright}
RESCEU--3/02  \\
KEK--TH--817   \\
TU--649       \\
hep--ph/0206248
\end{flushright}
\vspace{0.5cm}
\begin{center}
{\Large
{\bf 3- and 2-body supersymmetric processes:}                \\[1.1ex]
{\bf predictions, challenges and prospectives}}              \\[5ex]
 Francesca~Borzumati$^{\,a,b}$, Jae~Sik~Lee$^{\,b}$,
                     and Fumihiro~Takayama$^{\,c}$                 \\[1ex]
$^{a}${\it Research Center for the Early Universe,
          University of Tokyo, Tokyo 113-0033, Japan}        \\          
$^{b}${\it Theory Group, KEK, Tsukuba, Ibaraki 305-0801,
            Japan}                                           \\  
$^{c}${\it  Department of Physics, Tohoku University, 
          Sendai 980-8578,Japan}            
\end{center}
\vspace{1.5cm}
\begin{center} 
ABSTRACT \\
\vspace*{8mm}
\parbox{15cm}{The processes subject of this talk are: {\it 1)} the
$2\to 3$ processes contributing, together with the $2\to 2$ one, to
the hadron-collider production of a charged Higgs boson in association
with a $t$-quark; {\it 2)} the $2\to 2$ and $2\to 3$ processes giving
rise to the hadron-collider strahlung of a charged slepton from a
$t$-quark in $R_p$-violating models in which lepton-number-violating
trilinear couplings largely dominate over bilinear ones; {\it 3)}
$3$-body neutralino and chargino decays in similar $R_p$-violating
scenarios.  The significance of the $R_p$-violating processes is
critically assessed against implications from neutrino
physics. Contributions to neutrino masses arising at the tree level,
at the one- and two-loop levels are reviewed.  Comments are also made
on the production of sneutrinos via gluon fusion and on the decay of
sneutrinos into photons pairs, and on the influence that constraints
from neutrino physics have on these processes.
}
\end{center}

%\vspace*{4truecm}
\vfill 
{\begin{center} 
{\it Contributed by F. Borzumati to}                       \\
{\it Appi2002, Accelerator and Particle Physics Institute} \\
{\it Appi, Iwate, Japan, February 13--16 2002}
\end{center}}

\vfill
\end{titlepage}

\thispagestyle{empty}

%%%%%%%%%%%%%%%%%%%%%%%%%%%%%%%%%%%%%%%%

\begin{center}
{\Large
{\bf 3- and 2-body supersymmetric processes:}                 \\[1.1ex]
{\bf predictions, challenges and prospectives}}               \\[3ex]
{\bf  Francesca~Borzumati$^{\,a,b}$, Jae~Sik~Lee$^{\,b}$,
                     and Fumihiro~Takayama$^{\,c}$}           \\[2ex]
$^a$ {\it Research Center for the Early Universe,
          University of Tokyo, Tokyo 113-0033, Japan}         \\          
$^b$ {\it Theory Group, KEK, Tsukuba, Ibaraki 305-0801, Japan}\\  
$^c$ {\it  Department of Physics, Tohoku University, 
          Sendai 980-8578,Japan}                              \\[8ex]
\end{center}
{\begin{center} ABSTRACT \end{center}}  
\vspace{-4truemm}
{\small{The processes subject of this talk are: {\it 1)} the
$2\to 3$ processes contributing, together with the $2\to 2$ one, to
the hadron-collider production of a charged Higgs boson in association
with a $t$-quark; {\it 2)} the $2\to 2$ and $2\to 3$ processes giving
rise to the hadron-collider strahlung of a charged slepton from a
$t$-quark in $R_p$-violating models in which lepton-number-violating
trilinear couplings largely dominate over bilinear ones; {\it 3)}
$3$-body neutralino and chargino decays in similar $R_p$-violating
scenarios.  The significance of the $R_p$-violating processes is
critically assessed against implications from neutrino
physics. Contributions to neutrino masses arising at the tree level,
at the one- and two-loop levels are reviewed.  Comments are also made
on the production of sneutrinos via gluon fusion and on the decay of
sneutrinos into photons pairs, and on the influence that constraints
from neutrino physics have on these processes.
}}

\section{Introduction}
\label{intro} 
Now that LEP has been shut down, the upgraded Tevatron, with 
$\sqrt{s}=2\,$TeV and, possibly, a luminosity of $10-30$ fb$^{-1}$ 
by the end of the experiment~\cite{TEV}, and soon the LHC, with 
$\sqrt{s}=14\,$TeV and a luminosity of $100\,$fb$^{-1}$ per 
year~\cite{LHC}, will be the next machines to probe physics beyond the
Standard Model and to search for new particles.  Possible discoveries
will then be confirmed and thoroughly studied at hopefully incoming
Linear Colliders~\cite{LCs}.  The discovery reach of hadron machines,
however, may not be very high for particles that carry only weak
charges. Typically these particles are produced through the Drell-Yan
mechanism or through cascade decays.  At the LHC the former
mechanism suffers from being quark-initiated instead than
gluon-initiated. The latter is, in general, penalized at both machines
by the fact that the energy of the decaying particle may be
substantially reduced with respect to that of the initial beam.  The
discovery reach for sleptons, for example, is only $\ltap 350\,$GeV at
the LHC~\cite{SLEPTON}.  Nevertheless, a sufficient production cross
section may be achieved through the enhancement of some large
parameter, or if an alternative production mechanism is possible.

In the case of the charged Higgs there are two such large parameters:
$y_t$ and $\tan \beta$.  The alternative mechanism is the Higgs
strahlung, i.e. the production in association with a $t$- and a
$b$-quark~\cite{chtwotothreeA,BKP,chtwotothreeB,MR}, or in association
with a $t$-quark only~\cite{chtwototwo,BKP,MR}.  The former
possibility is described by the elementary $2 \to 3$ processes
$gg,\, qq \to t H^- \bar{b}$, the latter, by the $2 \to 2$
process $g b \to t H^-$.  Both types of elementary processes are
discussed in Section~\ref{chiggs}.  Predictions for the production
cross sections are also given, in the case in which the subsequent
decay of the charged Higgs gives rise to an overall multi-$b$'s
topology, as well as in the case in which at most two $b$-quarks are
present in the final state.  In the first case, the need of minimizing
the number of $b$-quarks to be tagged requires the combination of the
$2\to3$ and the $2\to 2$ cross sections.

The discovery of a charged Higgs boson will imply the existence of a
second Higgs doublet and, possibly, of a supersymmetric underlying
structure of the world~\cite{MSSM}.  Charged Higgs bosons give
potentially large contributions to flavour-changing-neutral-current
processes. Indeed, there are quite strong constraints on their mass
coming from these processes, in particular 
$b \to s \gamma$~\cite{LIMbsgsusy}. The constraint is even stronger in
a non-supersymmetric 2 Higgs Doublet Model~(2HDM)~\cite{LIMbsg2hdm1},
since it is not plagued by the model dependence that it has in
supersymmetric models.  However, it was shown that, even in this case,
this bound can be evaded if the 2HDM is the remnant of a multiHiggs
doublet model, with all doublets, except two, very
heavy~\cite{m2HDM}. The cross section for the strahlung production
mechanism, as well as the branching ratios for the decays of the
resulting charged Higgs are, in such a model, the same as in the 2HDM.
Therefore, it is reasonable to keep separated limits from virtual
effects and limits from direct production (as indeed it was/is already
assumed in direct searches at LEP and at the Tevatron). They are
complementary and they possibly reinforce each other.  We assume only
the current limits on the charged Higgs bosons mass coming from LEP
($\simeq 78.6\,$GeV) and the Tevatron ($\sim 160\,$GeV) (see for
example~\cite{lastLesHouches}), although these limits were challenged
as model dependent~\cite{BD}.

{}From charged-Higgs strahlung, the step to slepton strahlung in
$R_p$-violating models~\cite{RPV,RPV2} is quite small. If $R_p$ is
violated, or more properly, if the lepton number $L$ is violated, the
charged Higgs has all the quantum numbers of a charged slepton, say
$\tilde \tau$. This can then be singly produced in association with a
$t$- and a $b$-quark, or only a $t$-quark, exactly like the charged
Higgs. (Note that the single production of charged sleptons is
impossible if $R_p$ is conserved.)  The corresponding cross section
can be obtained generalizing the calculation for the associated
production of the charged Higgs.  The role of $y_b \cdot \tan \beta$
is plaid in this case by the $L$-violating parameter
$\lambda^\prime_{333}$, but there is no counterpart for the parameter
$y_t$.  The yield of $\tilde\tau$'s turns out to be considerable even
for modest values of $\lambda^\prime_{333}$.

The same production mechanism holds also for first and second
generation sleptons, $\tilde \mu$ and $\tilde e$.  The corresponding
cross sections are in these cases proportional to the square of
$\lambda^\prime_{233}$ and $\lambda^\prime_{133}$.  As in the case of
the charged Higgs, the identification of all decay modes for these
singly produced sleptons is very important in that it may affect the
way the production cross section must be calculated.  These issues are
discussed in Section~\ref{slstrahl}.  One of the slepton decay modes,
induced by the same $L$-violating coupling $\lambda^\prime_{i33}$
responsible for its production, $b \bar{t}$, is exactly one of the
decay modes of the charged Higgs. For this decay mode the combination
of the $2\to3$ and $2 \to 2$ elementary processes is mandatory.  The
other decay modes of the charged slepton are into a charged lepton and
the lightest neutralino, $\tilde{\ell}^- \to \ell^- \tilde{\chi}^0_1$,
or into a neutral lepton and the lightest chargino, 
$\tilde{\ell}^- \to \nu_{\ell} \tilde{\chi}^-_1$.  The relevance for
slepton production of other $R_p$-violating couplings involving at
least one third generation index, such as $\lambda^\prime_{i3j}$ and
$\lambda^\prime_{ij3}$ is also discussed.  Moreover, it is emphasized
that the simultaneous presence of couplings of type
$\lambda^\prime_{i33}$ and of type $\lambda_{i33}$ may have important
consequences for charged Higgs searches.

It is therefore crucial to know whether among the final states
obtained after the decays of $\tilde{\chi}^0_1$ and
$\tilde{\chi}^-_1$, there are any distinguishable from those obtained
from charged Higgs decays.  A discussion of the 3-body decays of
$\tilde{\chi}^0_1$ and $\tilde{\chi}^-_1$, due to the $R_p$-violating
coupling $\lambda^\prime_{i33}$, $\lambda^\prime_{332}$, and 
$\lambda^\prime_{323}$, is found in
Section~\ref{neuchardec}. It follows closely
Refs.~\cite{BGKT,CHARGINO} and it is rather general, i.e. valid for
any kinematical region for the lightest neutralino and chargino, and
independent of the assumption on gaugino mass unification at some
large scale.  Some guidelines on the distinguishability of a charged
slepton from a charged Higgs boson, both produced via strahlung from a
quark line, are given in Section~\ref{slstrahl}.

In the same spirit as in the discussion on charged Higgs boson
strahlung, also in the case of slepton strahlung and
neutralino/chargino decays, no constraints from virtual effects are
included in this analysis. As before, the rational is that these
constraints can, in general, be evaded by some (more or less) specific
choices of the different parameters contributing to these effects.  An
exception is made for neutrino physics in Section~\ref{sleptnuphys}.
The smallness of their masses, makes in general neutrino spectra very
sensitive to all $L$-number violating couplings.  We review in
Section~\ref{sleptnuphys} the main contributions to neutrino masses in
$R_p$-violating models, at the tree level, at the one- and at the
two-loop level.  The latter ones were recently analyzed in
Ref.~\cite{BL}.  In general, imposing that $R_p$-violating couplings
are compatible with neutrino physics jeopardizes the visibility of
some of the collider signals discussed here.  Moreover, the
constraints from the two-loop contributions, if those from the tree-
and one-loop level are evaded, may affect also searches for single
sneutrino $\tilde{\nu}$
production~\cite{BKP,Bar-Shalom:1998xz,Chaichian:2001vi} as well as
the decay $\tilde{\nu} \to \gamma \gamma$.  We discuss how these
constraints can still be evaded if, for example, $R_p$-violating
couplings have nontrivial phases. We single out the $R_p$-violating
parameters that remain unsuppressed after imposing constraints from
neutrino physics and the relevance that they may have not only for
slepton searches, but for Higgs searches as well. In any case, it is
clear that an independent collider probe of the lepton structure of
these couplings may become a precious tool for model building, in that
it may help individuating the real mechanism of generation of neutrino
masses.

Finally, we comment on the possibility of having very tiny violation
of the lepton number $L$. In general, it is assumed that this would
affect only the phenomenology of the lightest neutralino and possibly
neutrino physics.  We conclude our discussion in
Section~\ref{sleptnuphys}, by pointing out that in models in which
$\tilde{\chi}^0_1$ and $\tilde{\chi}^-_1$ are nearly degenerate, such
small violation of $R_p$ would have dramatic consequences also for the
phenomenology of the lightest chargino.

\section{Charged Higgs strahlung} 
\label{chiggs}

\begin{figure}[ht] 
\begin{center} 
\epsfxsize= 8.01cm 
\leavevmode 
\epsfbox{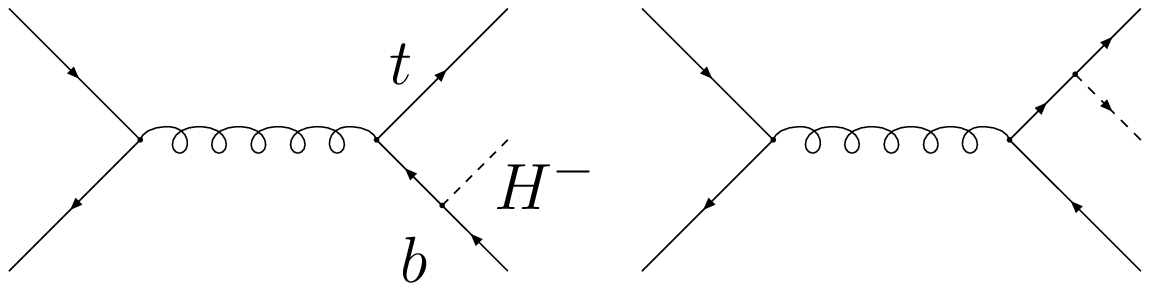} 
\end{center} 
\vspace*{-0.5truecm}
\caption[f1]{\small{Charged Higgs strahlung from the quark-initiated
 $2\to 3$ elementary process.}}
\label{qdiags} 
%\end{figure} 
%
%\begin{figure}[h] 
\vspace*{0.5truecm}
\begin{center} 
\epsfxsize= 10.35cm 
\leavevmode 
\epsfbox{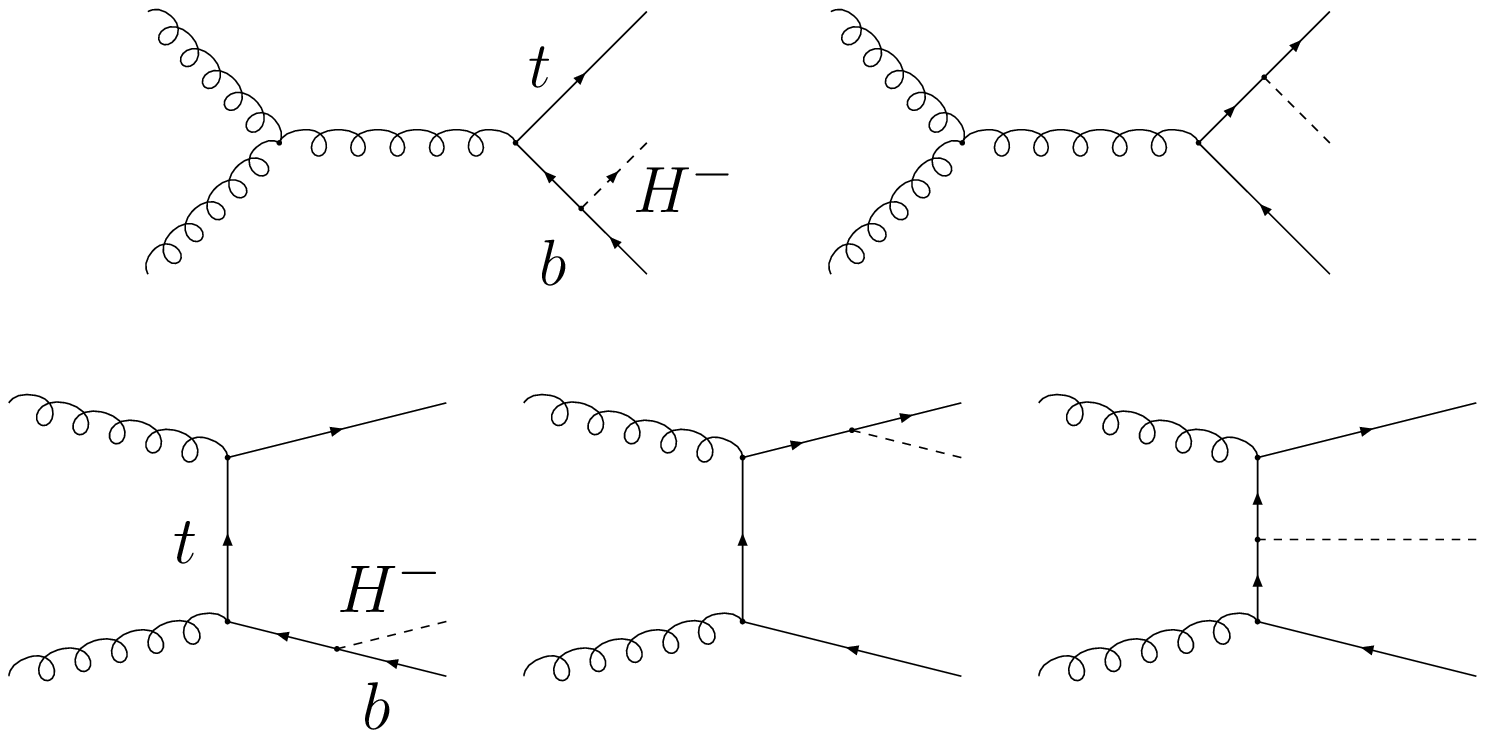} 
\end{center} 
\vspace*{-0.5truecm}
\caption[f1]{\small{Charged Higgs strahlung from the gluon-initiated
 $2\to 3$ elementary process.}}
\label{gdiags} 
\end{figure}

Charged Higgs strahlung is induced by the 
$2 \to 3$ elementary processes $gg,\,q\bar q\to t H^{-}\bar{b}$,  
which get contributions from the
Feynman diagrams in Figs.~\ref{qdiags} and~\ref{gdiags}.  The
production cross section for a $p \bar{p}$ ($p p$) collider is
obtained by convoluting the hard-scattering cross sections of the
quark- and gluon-initiated processes with the quark- and
gluon-distribution functions inside $p$ and/or $ \bar {p}$:
\begin{eqnarray}
 \sigma                  & = &  
 \frac{1}{2 s} \int^1_{\tau_{\rm min}} \frac{d \tau }{\tau}  
 \int^1_{\tau} \frac{d x}{x} \,
 \int d {\rm PS} (q_1+\!q_2;p_1,\!p_2,\!p_3) \times 
\nonumber \\             &   &          
\hspace*{1truecm}
 \left\{ \sum_{q}
 \Big[ q(x,\mu_f) \bar{q} (\tau/x,\mu_f) + q \leftrightarrow \bar{q}
 \Big] \vert {\cal M} \vert^2_{q \bar{q}}          \,  + \,
       g(x,\mu_f) g(\tau/x,\mu_f)  
       \vert {\cal M}\vert^2_{gg}  
 \right\} \,,                                         
\label{Xsec}
\end{eqnarray}
where $q_{i}$ ($p_{i}$) are the four-momenta of the initial partons
(final-state particles); 
$s$ is the hadron center-of-mass energy squared, whereas the parton
center-of-mass energy squared is indicated, as usual, by 
$\hat{s}= x_1 x_2\, s \equiv \tau s$; 
the functions $q(x,\mu_f)$,
$\bar{q}(x,\mu_f)$ and $g(x,\mu_f)$ designate respectively the quark-,
antiquark- and gluon-density functions with momentum fraction $x$ at
the factorization scale $\mu_f$, and the index $q$ in the sum, runs
over the five flavors $u, d, c, s, b$;
and $d$PS$(q_1+q_2;p_1, p_2, p_3)$ is an element of phase space 
of the 3-body final state.
The square amplitudes $\vert {\cal M} \vert^2_{gg}$ and 
$\vert {\cal M} \vert^2_{q \bar{q}}$ can be decomposed as:
\begin{eqnarray}
 \vert {\cal M} \vert^2_{q \bar{q}}  & = & 
 \left(4 \frac{G_F}{\sqrt{2}} M_W^2\right) \left( 4\pi \alpha_S\right)^2 
 \vert V_{tb}\vert^2 
 \left(v^2 \, V^{q \bar{q}}  + a^2 \, A^{q \bar{q}}\right)
\nonumber \\
 \vert {\cal M} \vert^2_{gg}         & = &
 \left(4 \frac{G_F}{\sqrt{2}} M_W^2\right) \left( 4\pi \alpha_S\right)^2 
 \vert V_{tb}\vert^2 
 \left( v^2 \, V^{gg} + a^2 \, A^{gg} \right)\,,
\label{amplitdecomp}
\end{eqnarray}
where color factors are included in the reduced squared amplitudes 
$V^{q \bar{q}}$, $ A^{q \bar{q}}$ and $V^{gg}$, $ A^{gg}$,
and where $v$ and $a$ are: 
\begin{equation}
 v  =   \frac{1}{2}
  \left(\frac{m_b}{M_W} \tan \! \beta + 
        \frac{m_t}{M_W} \cot \! \beta  \right)\,;   
\hspace*{1truecm}
 a  =   \frac{1}{2}
  \left(\frac{m_b}{M_W} \tan \! \beta - 
        \frac{m_t}{M_W} \cot \! \beta  \right) \,.
\label{vavalues}
\end{equation}
In the kinematical region $m_t > m_{H^\pm} +m_b$ the above cross
section could be well approximated by the much simpler resonant
production cross section, as given by the on-shell $t \bar t$
production cross section times the branching fraction for the
decay $\bar{t} \to H^- \bar{b}$. It was shown, however, that 
in the region $m_t \sim m_{H^\pm} + m_b$, this description 
fails to account for the correct mechanism of production and 
decay of the charged Higgs boson~\cite{SM}.

\begin{figure}[ht] 
\begin{center} 
\epsfxsize= 7.7cm 
\leavevmode 
\epsfbox{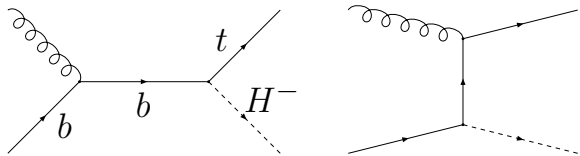} 
\end{center} 
\vspace*{-0.5truecm}
\caption[f1]{\small{Charged Higgs strahlung from the $2\to 2$ elementary
 process.}} 
\label{gbdiags} 
\end{figure}

The $2 \to 2$ process $g b \to t H^-$ giving also rise to 
strahlung of charged Higgs bosons has Feynman diagrams shown in 
Fig.~\ref{gbdiags}. The hard-scattering cross section is, in 
this case:
\begin{equation}
 \sigma (g\,b \to H^-\, t) = 
 \frac{1}{2 \hat{s}}
 \left(4\frac{G_F}{\sqrt{2}} M_W^2\right)  
 \left(4\pi\alpha_s\right)  \vert V_{tb}\vert^2 
 \left( v^2 V^{gb} + a^2 A^{gb} \right) 
\label{siggb}
\end{equation}
where the functions $V^{gb}$ and $A^{gb}$ include in this case the
reduced squared amplitudes as well as a phase-space integration
factor. Note that, since the initial $b$-quark is contained in the
proton or antiproton via a gluon, the $2\to 2$ process is actually of
the same order in $\alpha_s $ as the $2 \to 3$ ones.

When the decay of the charged Higgs does not give rise to 
%a pair of
third-generation quarks, the two production mechanisms are clearly
independent. They require the reconstruction of two $b$-quarks, in the
case of the $2\to 3$ processes, or of only one, in the case of the 
$2 \to 2$ process. This is the case of the channel 
$H^- \to \tau^- \nu_\tau$, suitable for the discovery of the charged
Higgs in the region of large $\tan \beta$, and possibly bound 
to play a key-role also in the region of intermediate $\tan \beta$ 
since it is not plagued by QCD background as the $H^-\to b \bar{t}$ 
mode~\cite{lastLesHouches}. The two production
mechanisms can be experimentally distinguished and
studied separately.

\begin{figure}[ht] 
\begin{center} 
\epsfxsize= 7.92cm 
\leavevmode 
\epsfbox{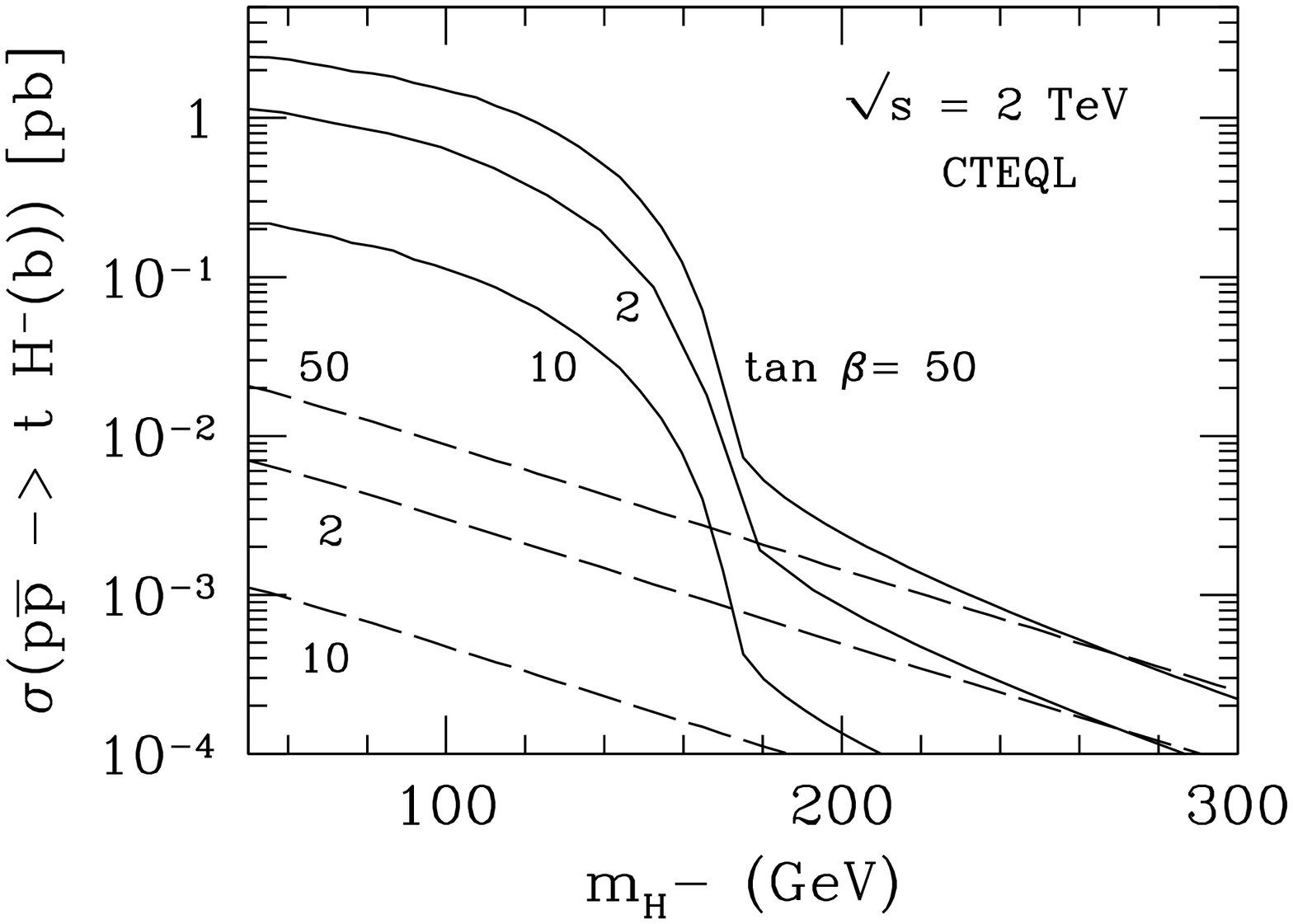} 
%\hspace*{0.4truecm}
\epsfxsize= 7.92cm 
\epsfbox{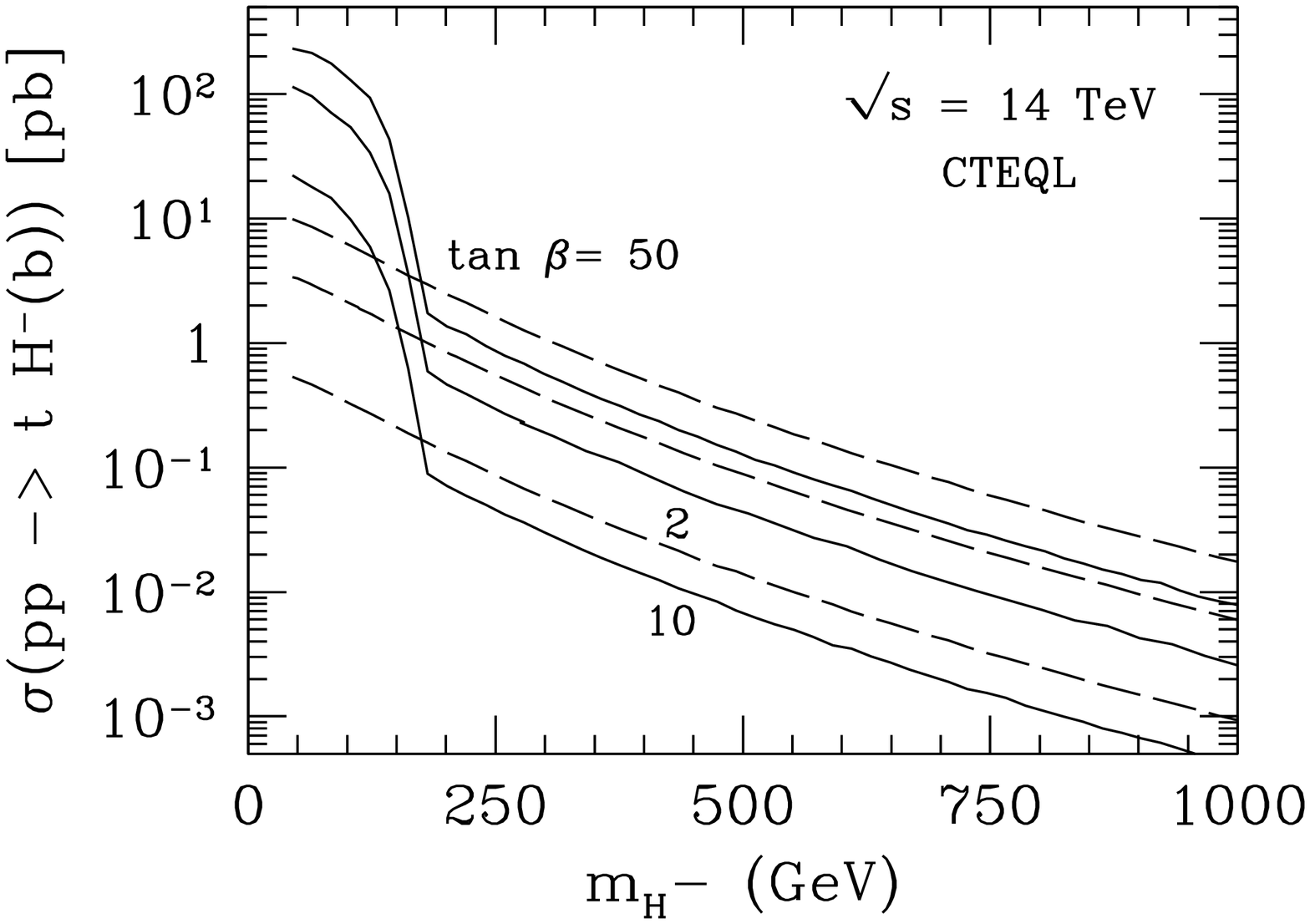} 
\end{center} 
\vspace*{-0.5truecm}
\caption[f1]{\small{Cross section 
 $\sigma(p\bar{p}\to t(\bar{b})H^- X)$ versus $m_{H^-}$, for
 $\sqrt{s} = 2\,$ and $14\,$TeV, for $\tan\!\beta = 2,\,10,\,50$. 
 The solid lines indicate the cross sections due to the $2\to 3$
 processes; the dashed lines, those due to the $2 \to 2$ process.
 Renormalization and factorization scales are fixed as
 $\mu_R\! =\! \mu_f\! =\! m_t+m_{H^-}$. From Ref.~\cite{BKP}.}}
\label{chindcrsec} 
\end{figure}

Leading-order predictions for the two production cross sections are
given in Fig.~\ref{chindcrsec} for the Tevatron
and LHC energies as a function of the charged-Higgs mass for three
different values of $\tan\beta$, $\tan \beta =2$, $10$, and $50$.  The
cross sections $\sigma(p \bar{p}, pp \to t H^- \bar{b} X)$ are shown
by the solid lines, those for $\sigma(p \bar{p}, pp \to t H^- X)$ by
dashed lines.  The leading-order parton distribution functions
CTEQ4L~\cite{CTEQ} are used, and the renormalization ($\mu_R$) and
factorization ($\mu_f$) scales are fixed to the threshold value
$m_t +m_{H^\pm}$.  A variation of these scales in the interval between
$(m_t+m_{H^\pm})/2$ and $2(m_t + m_{H^\pm})$ results in changes up to
$\pm 30\%$ in both cross sections.  QCD corrections, therefore, may be
important, as was shown in the case of associate production of the
neutral Higgs~\cite{NLOneutralstr}, but they are unfortunately not yet 
available. Part of these corrections is captured by the QCD
correction to the $b$-quark mass from which these cross sections 
depend. A study of their 
variation for different values of the $b$-quark mass can be found in 
Ref.~\cite{Higgs}.

As expected, the production cross section due to the $2\to 3$
processes is enhanced in the resonance region 
$m_{H^{\pm}} < m_{t}-m_{b}$, when $H^\pm$ is obtained as a decay
product of one of the two $t$-quarks produced on shell. The resonant
$t$-quark propagator is regularized by the width of the $t$-quark,
calculated from the SM decay $t\to b W^+$ and from the decay 
${t} \to H^{+}{b}$.  Away from the resonance region, the cross section
diminishes rather rapidly and becomes negligible at the Tevatron
energy. In this region, the relative
size of the two classes of cross section depends on $\sqrt{s}$ and
$m_{H^\pm}$.  Indeed, at high energies, where the gluon-initiated 
$2\to 3$ processes dominate over the quark-initiated ones, the cross
section arising from the elementary process $g b \to t H^-$ is larger
than that from $gg \to t H^- b$, which is penalized by a 3-body phase
space suppression.  At the Tevatron center-of-mass energy, the
quark-initiated $2 \to 3$ processes still have the dominant role up 
to intermediate values of $m_{H^\pm}$.  For the particular choice of
scales $\mu_R$ and $\mu_f$ made here, the cross-over for the two 
cross sections is at about $m_{H^\pm} \sim 265\,$GeV.
Both classes of cross sections show the typical behaviour as a 
function of $\tan \beta$, with a minimum at around 
$(m_t/m_b)^{1/2}$.

\begin{figure}[ht] 
\vspace*{0.3truecm}
\begin{center} 
\epsfxsize= 10.2cm 
\leavevmode 
\epsfbox{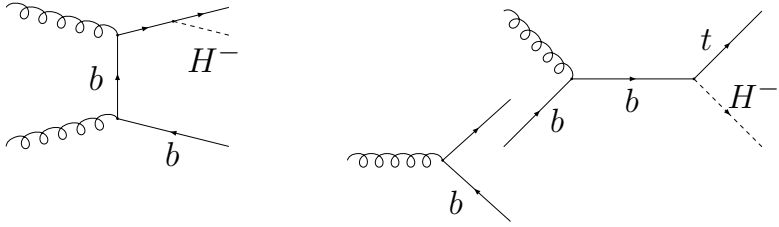} 
\end{center} 
\vspace*{-0.5truecm}
\caption[f1]{\small{Term of the gluon-initiated $2 \to 3$ process
 overlapping with a contribution from the $2 \to 2$ one.}} 
\label{overlap} 
\end{figure}

When the charged Higgs, decaying, gives rise to at least two
$b$-quarks, the final state to be identified contains at least three
$b$'s for the $2 \to 2$ production mechanism, and at least four for
the $2\to 3$ one.  This is the case of the decay channel 
$H^+ \to t \bar{b}$, particularly important in the region of small 
$\tan \beta$, when the rate for $H^-\to \tau^- \nu_\tau$ vanishes. 
Since tagging more than three $b$'s seems quite difficult 
even at the LHC (see, however, Ref.~\cite{FOURbTAG}), the two 
production mechanisms 
result into final states that are indistinguishable and a sum of the two
cross sections is necessary. As shown in
Fig.~\ref{overlap}, the kinematical region in which one of the two
initial gluons in the $2 \to 3$ processes produces a pair $b \bar{b}$
collinear to the initial $p$ or $\bar{p}$, is, by definition, also
kept into account by the $2\to 2$ process with the initial
$b$-quark collinear to the colliding $p$ or $\bar{p}$, in which it is
contained via a gluon.  A large factor
$\alpha_s(\mu_R)\log(\mu_f/m_b)$ ($\mu_f$ is ${\cal{O}}(m_{H^\pm})$),
is induced in this kinematical region in the cross section from the
$2\to 3$ processes. This same factor is contained in the cross section
from the $2 \to 2$ process and it is resummed to all orders,
$(\alpha_s(\mu_R) \log(\mu_f/m_b))^n$, when making use of the
phenomenological $b$-distribution function.  The first order, $n=1$,
must then be subtracted from the sum of the two cross
sections~\cite{BKP}.

\begin{figure}[ht] 
\begin{center} 
\epsfxsize= 7.92cm 
\leavevmode 
\epsfbox{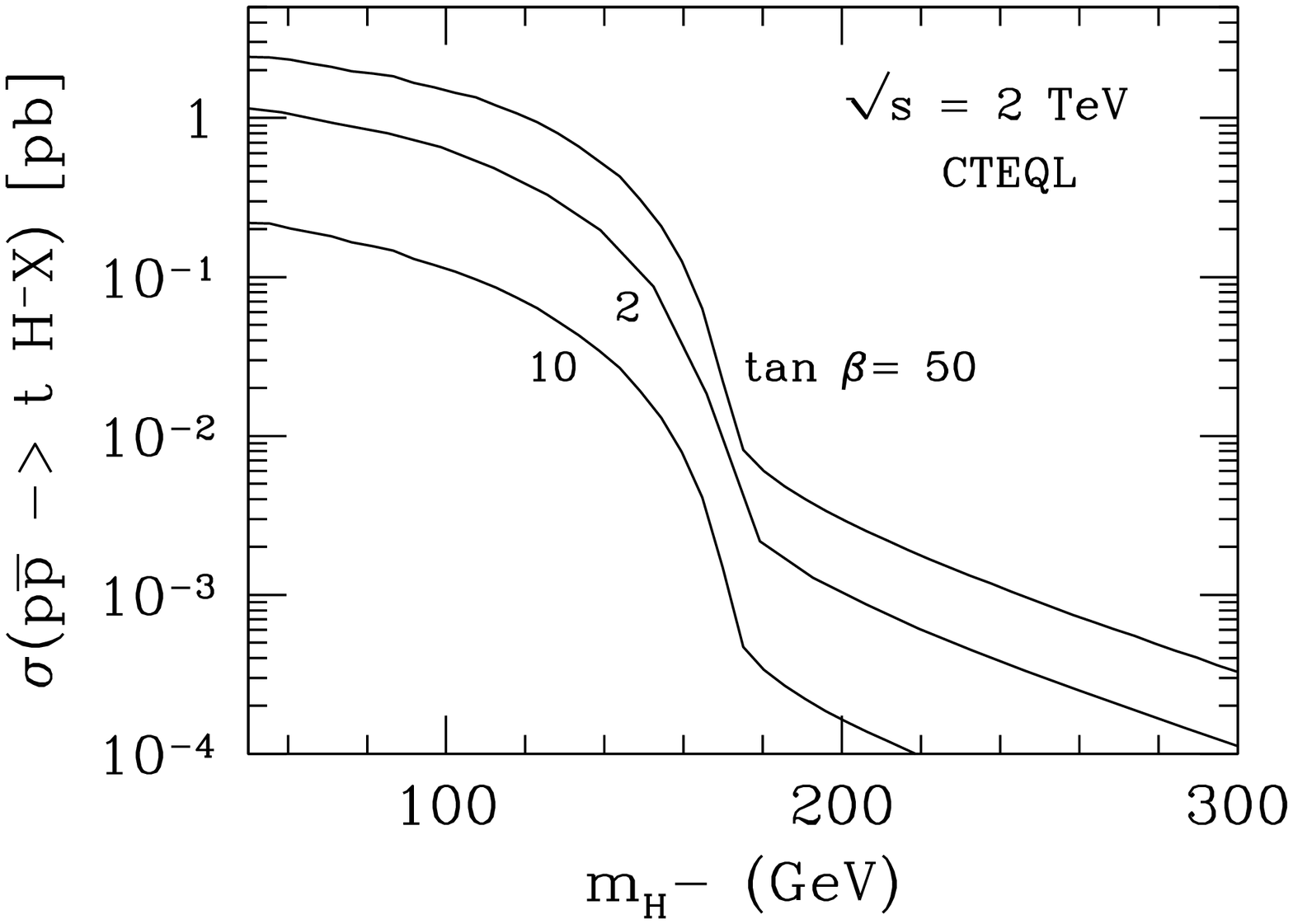} 
%\hspace*{0.4truecm}
\epsfxsize= 7.92cm 
\epsfbox{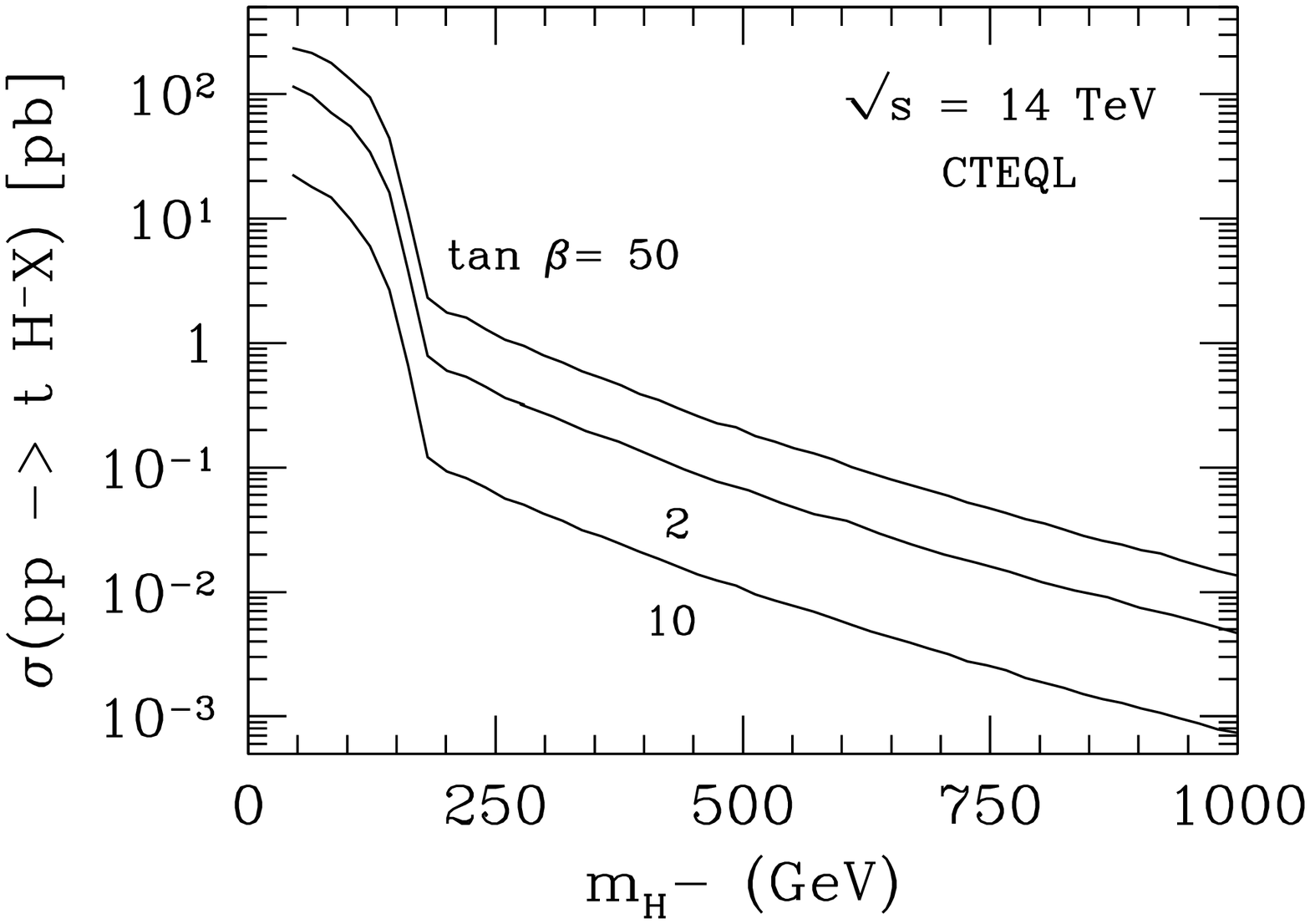} 
\end{center} 
\vspace*{-0.5truecm}
\caption[f1]{\small{The cross section $\sigma(p\bar{p}\to t H^- X)$
 obtained by adding the contributions from the $2 \to 2$ and $2 \to 3$
 processes shown in Fig.~\ref{chindcrsec}, and subtracting
 overlapping terms. From Ref.~\cite{BKP}.}} 
\label{chsumcrsec} 
\end{figure}

Predictions for the appropriately summed inclusive cross section 
are shown in Fig.~\ref{chsumcrsec}, both for the 
Tevatron and the LHC.  These cross sections have the same theoretical
uncertainty of the individual cross sections of Fig.~\ref{chindcrsec}, 
as well as the same $\tan \beta$ dependence. Once again, the 
inclusion of QCD corrections is awaited for.  (The dependence 
of these cross sections on the value of $m_b$, which may be 
most strongly affected by QCD corrections, is shown in 
Ref.~\cite{Higgs}.) Similarly, at least in 
the region of large $\tan \beta$, supersymmetric corrections should be 
nonnegligible. These have been calculated for the $2 \to 3$ processes, 
see~\cite{lastSOLA} and references therein. Recently also a calculation
of the gluino corrections for the 
$2 \to 2$ process~\cite{Gao:2002is} had appeared, but no complete
estimate exists, as yet, for the summed cross sections.

\section{Charged slepton strahlung} 
\label{slstrahl} 
As the charged Higgs boson, a charged slepton can be radiated from a
$t$-quark thanks to the couplings $\lambda^\prime_{i3j}$ in the
superpotential 
\begin{equation}
  W  \supset  - \lambda^\prime_{lmn}L_{l} Q_{m} D^c_{n} 
  - \frac{1}{2} \lambda_{lmn}       L_{l} L_{m} E^c_{n} \,, 
\label{superpot}
\end{equation}
where, for completeness also the operators with couplings 
$\lambda_{lmn}$ are shown.  
In particular, for the coupling $\lambda^\prime_{333}$, both the 
$2\to 3$ partonic processes, 
$q \bar{q}, gg \to t \bar{b} \tilde{\tau}$ and the $2 \to 2$ process
$g b \to t \tilde{\tau}$ contribute to the strahlung of a
$\tilde{\tau}$-slepton.  The corresponding Feynman diagrams are
obtained from those in Figs.~\ref{qdiags}, \ref{gdiags},
and~\ref{gbdiags} by replacing $H^-$ with $\tilde{\tau}$.  The
inclusive cross section induced by the $2 \to 3$ processes is that of
Eq.~(\ref{Xsec}) with the square amplitudes 
$\vert {\cal M} \vert^2_{gg}$ and $\vert {\cal M} \vert^2_{q \bar{q}}$
given by:
\begin{eqnarray}
 \vert {\cal M} \vert^2_{q \bar{q}}  & = & 
 \frac{1}{4} \vert\lambda^{\prime}_{333}\vert^2 
 \left( 4\pi \alpha_S\right)^2  \vert V_{tb}\vert^2 
 \left( V^{q \bar{q}}  +  A^{q \bar{q}}\right)
\nonumber \\
 \vert {\cal M} \vert^2_{gg}         & = &
 \frac{1}{4} \vert\lambda^{\prime}_{333}\vert^2
 \left( 4\pi \alpha_S\right)^2  \vert V_{tb}\vert^2 
 \left( V^{gg} +  A^{gg} \right)\,,
\label{rpamplitdecomp}
\end{eqnarray}
where the values $v=a=1/2$ were already substituted in.  The reduced
square amplitudes $V^{q \bar{q}}$, $ A^{q \bar{q}}$ and $V^{gg}$, $
A^{gg}$ coincide with those obtained for the $2 \to 3$
charged-Higgs-production processes, once the replacement 
$m_{H^\pm} \to m_{\tilde{\tau}}$ is made.  The partonic cross section
for $g b \to t \tilde{\tau}$ can also be obtained from that for
charged Higgs production, with obvious changes, i.e.  by replacing 
$4 {G_F} M_W^2 /{\sqrt{2}}$ with $\vert \lambda^\prime_{333}\vert^2$
and $m_{H^-}$ with $m_{\tilde{\tau}}$. 

\begin{figure}[ht] 
\begin{center} 
\vspace*{0.3truecm}
\epsfxsize= 7.92cm 
\leavevmode 
\epsfbox{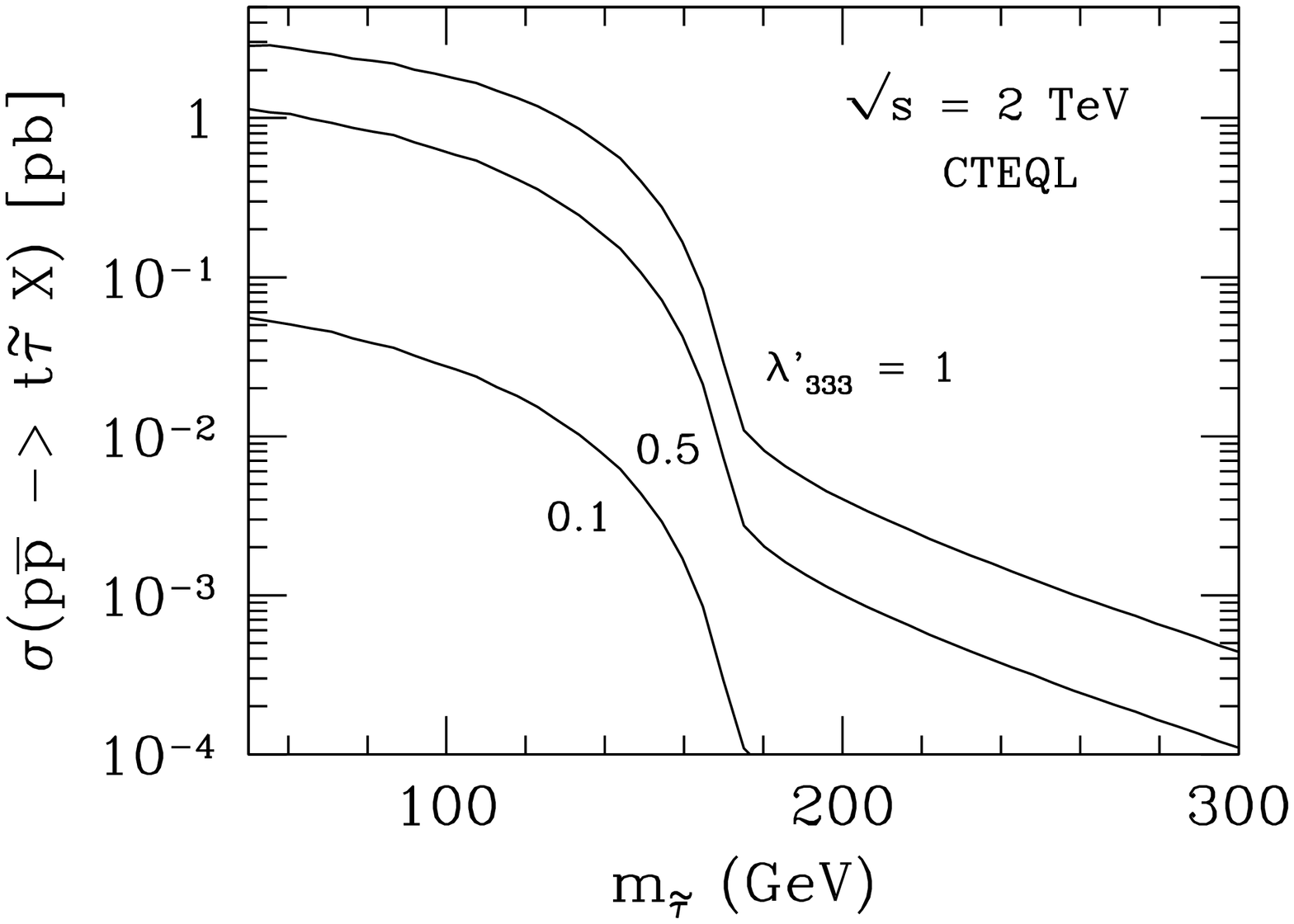} 
%\hspace*{0.4truecm}
\epsfxsize= 7.92cm 
\epsfbox{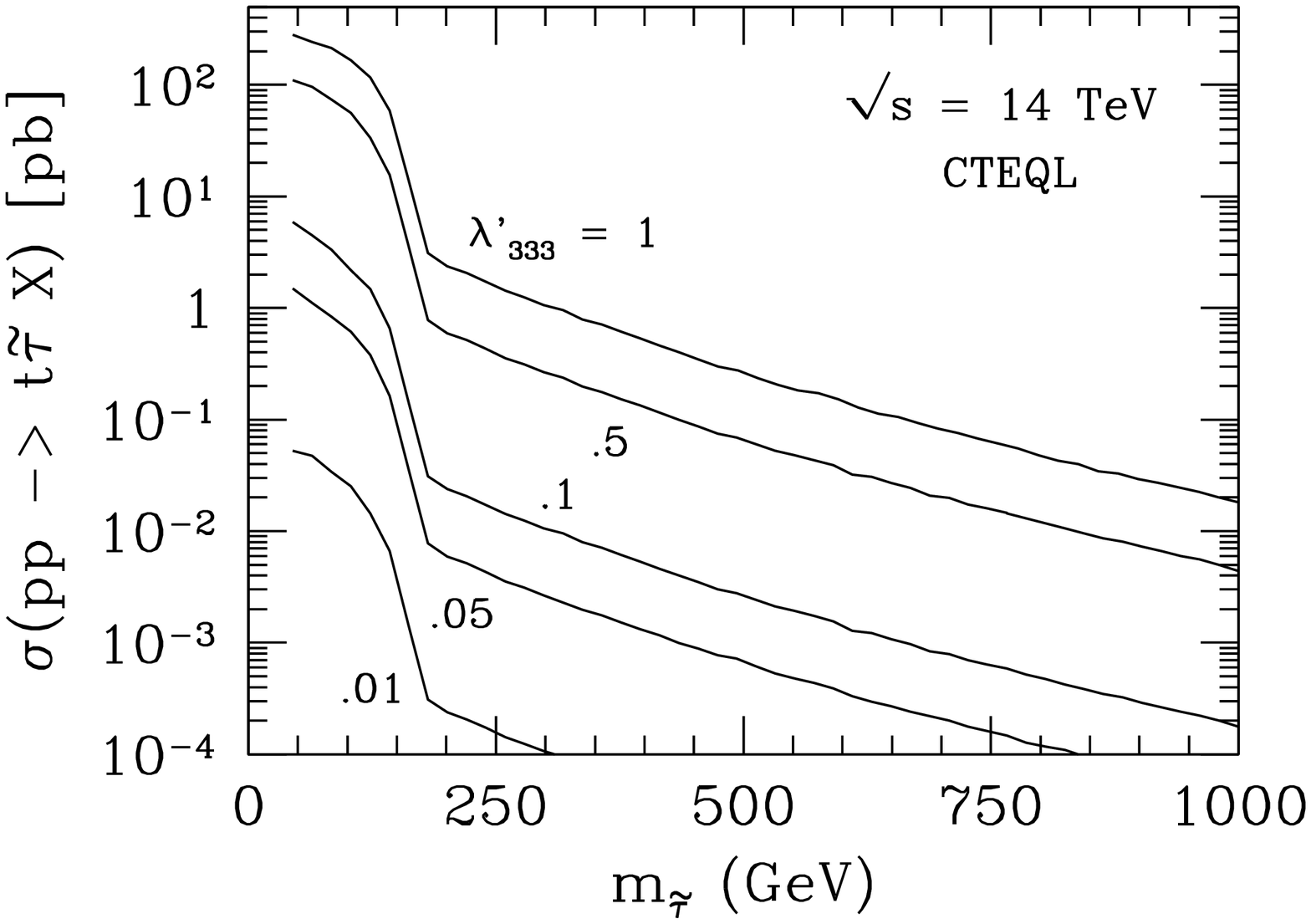} 
\end{center} 
\vspace*{-0.5truecm}
\caption[f1]{\small{Same as in Fig.~\ref{chsumcrsec}, for 
 $\tilde{\tau}$ production, when the coupling $\lambda^\prime_{333}$
 is dominant among all other $R_p$-violating couplings.  The
 production cross section is shown for different values of 
 $\lambda^\prime_{333}$. From Ref.~\cite{BKP}.}} 
\label{stausumcrsec} 
\end{figure}

The same coupling $\lambda^\prime_{333}$ inducing the radiation of the
charged slepton from a third generation quark line, is also
responsible for the decay $\tilde{\tau}\to b \bar{t}$.  In this case,
the charged slepton behaves exactly like a charged Higgs decaying into
the same final state, and has only the effect of increasing the number
of $t\bar{t} b\bar{b}$ events from $2\to 3$ processes, or of $t
\bar{t} b$ events from the $2 \to 2$ one.  In this case, the
evaluation of the slepton production cross section requires the
combination of the $2\to 3$ and the $2\to 2$ cross sections, as
outlined in Section~\ref{chiggs}. Predictions for this cross section
are shown in Fig.~\ref{stausumcrsec} for the Tevatron and the LHC,
where in the kinematical region of a resonant $t$-quark, only the two
decay modes $t\to W^+ b $ and ${t} \to \tilde{\tau}^{+}{b}$ are
considered. It is assumed that the charged Higgs $H^\pm$ is
sufficiently heavy, as to kinematically forbid the decay mode
${t}\to H^+ {b} $.

The large cross section obtained in the case of $\sqrt{s} = 14\,$TeV
implies that at the LHC, with a luminosity of 100 fb$^{-1}$ per year,
light $\tilde{\tau}$'s may be produced in abundance even for couplings
as small as 0.01. Charged sleptons are, therefore, potentially more
accessible (for not too small values of $\lambda^\prime_{333}$) than
in models with lepton-number conservation, where their direct
production relies on the Drell--Yan process. As it is known, this
allows a discovery reach for sleptons only up to 
$m_{\tilde{l}} \ltap 350\,$GeV at the LHC~\cite{SLEPTON}.  Cascade
decays may provide the bulk of slepton production in the
lepton-number-conserving models, but such processes are highly
model-dependent. They also complement slepton production in the
lepton-number-violating case studied here.  The potential reach at the
Tevatron is limited to large couplings and/or light $\tilde{\tau}$'s,
{\it i.e.} to $\tilde{\tau}$ lighter than the
$t$-quark~\cite{TOPsearches}.  In the resonant part of the curves,
where the cross section can be approximated by the production cross
section of a pair of $t$-quarks times the branching ratio for the
decay of one of the two $t$'s, the direct decay 
$t\to \tilde{\tau}^+b$ can significantly constrain the values of
$\lambda^\prime_{333}$~\cite{HANMAGRO}. For 
$m_{\tilde{\tau}}= 70\,$GeV, only values up to $\sim 0.2$ will remain
allowed, if charged slepton searches have a negative outcome at Run~II
with an integrated luminosity of $10\,$fb$^{-1}$~\footnote{The
 analysis of Ref.~\cite{HANMAGRO} neglects the decay
 $\tilde{\tau}\to \nu_\tau \tilde{\chi}^-_1$, with the subsequent
 chargino decay $\tilde{\chi}^-_1 \to b \bar{b}\tau^-$, which may be
 nonnegligible for the values of $\lambda^\prime_{333}$ considered}.

Besides the decay $\tilde{\tau} \to \bar{t} b$, two additional 
2-body decays, due to gauge interactions, are possible for 
$\tilde{\tau}$: 
$\tilde{\tau} \to \tilde{\chi}^{0}\tau^-$ and 
$\tilde{\tau} \to \tilde{\chi}^{-}\nu_{\tau}$.
Whether the $2\to 3$ processes and the $2 \to 2$ one must be combined
also in this case, depends on the final states produced by the decays
of charginos and neutralinos.  A list of all the final states produced
by the couplings $\lambda^\prime_{i33}$ and $\lambda_{i33}$ can be
found in Table~\ref{Table1}, in which also the final states obtained
from charged Higgs decays are listed.  In first approximation, let us
assume that the coupling $\lambda^\prime_{333}$ is dominant among all
$R_p$-violating couplings. In such a case, the lightest neutralino
decays as (see Section~\ref{neuchardec}):
\begin{equation}
 \tilde{\chi}^0_1 \,\to\, b \bar{b} \nu_\tau  \,, \quad \quad 
 \tilde{\chi}^0_1 \,\to\, t \bar{b} \tau^-    \,, \quad \quad 
 \tilde{\chi}^0_1 \,\to\, \bar{t} b \tau^+    \,, 
\label{neutdecRP}
\end{equation}
with the second and third decay mode allowed only for a heavy
$\tilde{\chi}^0_1$, i.e. for 
$m_{\tilde{\chi}^0_1} > m_t + m_b +m_\tau$.  The 2-body decays due to
tree-level mixing of Higgs--slepton, $\tilde{\chi}^{-}$--$\tau^-$, and
$\tilde{\chi}^{0}$--$\nu_{\tau}$ are neglected here. They become
competitive with the $3$-body decays only when the values of the
trilinear couplings are rather small, and therefore no slepton
strahlung signal is expected. For a discussion of the decays due to
these mixings, see for example Ref.~\cite{CHUNneutr}.  In the same
approximation, the lightest chargino decays via gauge interactions as:
\begin{equation}
 \tilde{\chi}^-_1 \,\to\, W^- \tilde{\chi}^0_1\,. 
\label{chardecGAUGE}
\end{equation}
The $R_p$-violating chargino decays,
\begin{equation}
 \tilde{\chi}^-_1 \,\to\, b \bar{b} \tau^-   \,, \quad \quad 
 \tilde{\chi}^-_1 \,\to\, b \bar{t} \nu_\tau \,, \quad \quad 
 \tilde{\chi}^-_1 \,\to\, t \bar{t} \tau^-   \,, 
\label{chardecRP}
\end{equation}
with the second and third decays allowed only if 
$m_{\tilde{\chi}^-_1} > m_t + m_b$ and 
$m_{\tilde{\chi}^-_1} > 2 m_t + m_\tau$, respectively, are competitive
with the decay mediated by gauge interactions only if
$\lambda^\prime_{333}$ is relatively large.  The possible final states
that can be produced through these decays contain all at least 3
$b$'s. The $2\to 3$ and $2 \to 2$ processes should, then, be combined,
if the number of $b$-quarks that can be detected has to be minimized.
\begin{table}[ht]
\begin{center}
\vskip 0.5cm
\begin{tabular}{|c||l||l|l|}
\hline 
&&\multicolumn{2}{c |} {}\\
$H^-$/$\tilde{\ell}_i^-$ &\quad Decay mode  \quad 
                         &\multicolumn{2}{c |} {Decay final state \quad}\\  
&&\multicolumn{2}{c |}{}\\
\hline\hline
&& &\\
         &\quad $b\bar{t} $      &\quad $b(\bar{b} \ell_i^- \nu_i)$
         & \\[1.001ex]
         &\quad $\tau^-\nu_\tau$ & &\quad $\tau^-\nu_\tau$\\[1.001ex]
$H^-$    &\quad $hW^- $      &\quad $(b\bar{b})(\ell_i^- \nu_i)$\quad
         &\quad $(\tau^- \tau^+) (\ell_i^- \nu_i)$ \quad \\[1.1ex]
\cline {3-4}
&&\multicolumn{2}{c |}{}\\[-0.999ex]
         &\quad $\tilde{\chi}^-\tilde{\chi}^0$ 
         &\multicolumn{2}{c |} 
              {\quad $W^- \tilde{\chi}^0 \tilde{\chi}^0$\quad} \\
&&\multicolumn{2}{c |}{}\\
\hline\hline 
&& &\\
         & \quad $b\bar{t} $ &\quad $b(\bar{b} \ell_i^- \nu_i)$ 
         &  \\[1.001ex]
         & \quad $\tau^- \nu_\tau $  & 
         & \quad $\tau^- \nu_\tau$   \\
&& &\\
\cline {2-4}
&& &\\
         & &\quad $\ell_i^-(b\bar{b}\nu_i)$  
           &\quad $\ell_i^-(\tau^- \tau^+\nu_i)$  \\[1.001ex]
$\tilde{\ell}_i^-$ 
         & \quad $\ell_i^- \tilde{\chi}^0$ 
           &\quad $\ell_i^-(t\bar{b}\ell_i^-)$ 
           &\quad $\ell_i^-(\nu_\tau \tau^+\ell_i^-)$\\[1.001ex]
         & &\quad $\ell_i^-(\bar{t}b \ell_i^+)$ 
           &\quad $\ell_i^-(\nu_\tau \tau^- \ell_i^+)$ \\
&& &\\
\cline {2-4}
&& &\\
         & &\quad $\nu_i(b\bar{b}\ell_i^-)$ 
           &\quad $\nu_i(\tau^- \tau^+\ell_i^-)$\\[1.001ex]
         &\quad $\nu_i \tilde{\chi}^-$ 
           &\quad $\nu_i(b\bar{t}\nu_i)$ 
           &\quad $\nu_i(\tau^- \nu_\tau\nu_i)$\\[1.001ex]
         & &\quad $\nu_i(t\bar{t}\ell_i^-)$ 
           &\quad $\nu_i(\nu_\tau \nu_\tau \ell_i^-)$\\[1.005ex]
         & &\quad $\nu_i\tilde{\chi}^0W^-$ 
           &\quad $\nu_i\tilde{\chi}^0W^-$ \\
&& &\\
\hline 
\end{tabular}
\end{center}
\caption[]{\small{Comparison between decay modes and final states obtained
from a charged Higgs boson in a $R_p$-conserving scenario and those
obtained from a charged slepton of generation $i$ in a $R_p$-violating 
one, in which the couplings $\lambda^\prime_{i33}$ and 
$\lambda_{i33}, (i\ne3)$, are the dominant one.}} 
\label{Table1}
\end{table}

In this case, distinguishing a light $\tilde{\tau}$ (with also light
neutral gauginos) from a charged Higgs will be difficult. The final
states from $\tilde{\tau}$ production and decay are very similar to
those obtained from charged Higgs bosons: compare for example $t
(\bar{b}) b \bar{b} \tau^- \nu_\tau X$ obtained from sleptons, versus
the states $t (\bar{b}) \bar{t} b X$, $t (\bar{b}) \tau^- \nu_\tau X$,
obtained from the charged Higgs boson decays into $\bar{t}b$ and
$\tau^- \nu_\tau$, or versus the state $t (\bar{b}) b\bar{b} \tau^-
\nu_\tau X$ obtained, for example, from the decay of the charged Higgs
boson into $h W^-$.

For heavier sleptons and at least one of the two neutral gauginos
heavier than the $t$-quark, it is possible  
to obtain the interesting final state $(2t) \bar{b} (2\tau^-) X$.  When
the two $t$'s decay hadronically, the resulting signature is that of
two equal-sign leptons, jets and no missing energy.  Note that the two
equal-sign leptons may be two $\mu$'s if the coupling
$\lambda^\prime_{233}$ is the one considered. In this case, the
strahlung-produced sleptons are $\tilde{\mu}$'s. The production cross
sections shown in Fig.~\ref{stausumcrsec} apply also to this case,
when $\lambda^\prime_{333}$ is substituted by $\lambda^\prime_{233}$
and $m_{\tilde{\tau}}$ by $m_{\tilde{\mu}}$.

\begin{figure}[ht] 
\begin{center} 
\epsfxsize= 7.7cm 
\leavevmode 
\epsfbox{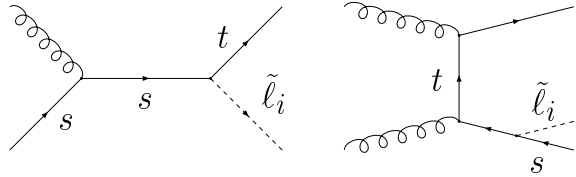} 
\end{center} 
\vspace*{-0.5truecm}
\caption[f1]{\small{Strahlung of a charged slepton associated with a 
 $t$- and an $s$-quark.}} 
\label{tsdiags} 
\end{figure}

The couplings $\lambda^\prime_{i3j}$ give also rise to strahlung of
the slepton $\tilde{\ell}_i$ in association with a $t$- and an
$s$-quark, when $j=2$, or in association with a $t$- and a $d$-quark
when $j=1$.  At the partonic level these production mechanisms are
described by the elementary processes
$q \bar{q}, gg \to t \bar{s} \tilde{\ell}_{i}$,
$g s \to t \tilde{\ell}_{i}$, and  
$q \bar{q}, gg \to t \bar{d} \tilde{\ell}_{i}$,
$g d \to t \tilde{\ell}_{i}$.  Some representative diagrams for the
strahlung associated with a $t$- and an $s$-quarks are shown in
Fig.~\ref{tsdiags}.  For these couplings, at most 2 $b$'s are present
in the final states and they can be easily tagged at the LHC. The 
$2 \to 3$ and $2\to 2$ processes can therefore be disentangled and
studied separately.

Note that the cross sections for the $2 \to 3$ processes 
$q \bar{q}, gg \to t \bar{s} \tilde{\ell}_{i}$, and 
$q \bar{q}, gg \to t \bar{d} \tilde{\ell}_{i}$ are as those obtained
for $q \bar{q}, gg \to t \bar{b} \tilde{\tau}$.  The states 
$t\bar{t} s \bar{s}$ and $t\bar{t} d \bar{d}$, obtained when the
radiated sleptons decay as $\tilde{\ell}_{i} \to s \bar{t}$ and
$\tilde{\ell}_{i} \to d \bar{t}$, cannot be mimicked by charged Higgs
bosons production and decays, which are penalized by CKM suppression
and by the practically vanishing mass of the $s$- and $d$-quark.

The production through the $2\to 2$ processes 
$g s \to t \tilde{\ell}_{i}$ and $g d \to t \tilde{\ell}_{i}$ (in 
particular also $\tilde{\tau}$ production) is larger than the 
production through $g b \to t \tilde{\ell}_{i}$ because of the 
larger content of the $s$- and $d$-quarks in the proton.  The 
possibility of distinguishing 
which coupling has induced the production of $\tilde{\ell}_i$, i.e. 
whether $\lambda^\prime_{i33}$, $\lambda^\prime_{i32}$, or 
$\lambda^\prime_{i31}$, strongly depends on how accurately the final
states can be studied.  The final states induced by the 
coupling $\lambda^\prime_{i3j}$ through $2\to 2$ processes are:
$$
\begin{array}{ll}
 t \bar{t} d_j                  \quad \quad 
               & (\tilde{\ell}_i \to d_j \bar{t})   
\\ 
 t \ell_i^- b \bar{d}_j \nu_i   \quad \quad 
               & (\tilde{\ell}_i \to \ell_i^- \tilde{\chi}^0_1 
                  \to        \ell_i^- b \bar{d}_j \nu_i \, /\,
                  \tilde{\ell}_i \to \nu_i \tilde{\chi}^-_1 
                  \to        \nu_i  b \bar{d}_j \ell_i^-)      
\\ 
 2 (t \ell_i^-) \bar{d}_j       \quad \quad 
               & (\tilde{\ell}_i \to \ell_i^- \tilde{\chi}^0_1 
                  \to        \ell_i^- t \bar{d}_j \ell_i^-)    
\\ 
 t \bar{t} d_j \nu_i \nu_i        \quad \quad 
               & (\tilde{\ell}_i \to \nu_i \tilde{\chi}^-_1 
                  \to        \nu_i \bar{t} d_j \nu_i)\,.
\end{array}
$$
(Chargino and neutralino decays due to the couplings
$\lambda^\prime_{i32}$ are listed in the next Section.)  The
identification of 3 $b$'s versus that of 2 $b$'s and one $s$-quark,
for example, will allow to distinguish between $\lambda^\prime_{i33}$
and $\lambda^\prime_{i32}$.  Also in this case, the final states
obtained when $j=1$ or $j=2$ are not likely to be induced by charged
Higgs bosons and such signals would be genuine $R_p$-violating ones.

There is another rather interesting possibility that opens up for
sleptons obtained from $2 \to 2$ processes, through the couplings
$\lambda^\prime_{i32}$ and $\lambda^\prime_{i31}$. If the sleptons 
$\tilde{\ell}_i$ produced in this way decay through the couplings 
$\lambda_{ij3}$ or $\lambda_{j3i}$, the processes 
\begin{equation}
 g \bar{d}, g \bar{s} 
      \to t \tilde{\ell}_i \to t \tau^- \nu_j 
\end{equation}
lead to the same final state obtained from charged Higgs bosons, also
originating from the $2 \to 2$ process $g \bar{b} \to t H^-$.  Given
the larger yield of the $s$- and the $d$-quark into the proton, the
slepton production cross section may be much larger than that of the
charged Higgs and a serious contamination of the charged Higgs signal
may be expected even for not too large values of the relevant
$R_p$-violating couplings.  Such a possibility should be kept in mind
when searching for charged Higgs bosons, through the mode 
$\tau^- \nu_\tau$. This is especially so, when this decay mode is used
to extend the region of $\tan \beta$ that can be probed to
intermediate values such as $\sim 10$, for which the Yukawa coupling
of the $\tau$ is not particularly large.
%Such contamination 
%are also to be expected when the sleptons decay through 
%the couplings $\lambda_{ij3}$ and $\lambda_{j3i}$. 

Finally, the coupling $\lambda^\prime_{ij3}$ 
give rise to slepton strahlung in association with a $c$- and a 
$b$-quark or with a $u$- and a $b$-quark, when $j=2$ and $j=1$. 
In this case, the strahlung mechanism must compete with the 
$2\to 1$ processes, $b \bar{c} \to \tilde{\ell}_i$ and 
$b \bar{u} \to \tilde{\ell}_i$, which lead to final states 
different from those obtained from the $2\to 2$ and $2\to 3$ 
processes. Among these, the states $c\bar{c} b \bar{b}$ may 
substantially add up to those obtained from $W$-pair production.  
In addition, final states such as $(2c) \bar{b} (2 \ell_{i})$, 
with two like-sign leptons, jets and no missing energy can be 
obtained, in this case, 
even for light neutralinos, and are distinctive of $R_p$-violating
couplings.

\section{Neutralino and chargino decays} 
\label{neuchardec} 
As already mentioned in Section~\ref{slstrahl}, in $R_p$-violating
scenarios with a dominant coupling $\lambda^\prime_{333}$, the
lightest neutralino decays as 
$\tilde{\chi}^0_1 \,\to\, b \bar{b} \nu_\tau$, $t \bar{b} \tau^-$, 
and $\bar{t} b \tau^+$.  A detailed analysis of these and other
decays was performed in Ref.~\cite{BGKT}, where no 
assumption was made as to whether the final state is massive
or massless, or whether gaugino masses are unified or not at some
large scale. As already mentioned in the previous Section, 
we neglect here the $2$-body decays of the lightest neutralino 
due to tree-level mixing Higgs--slepton,
$\tilde{\chi}^{-}-\tau$, and $\tilde{\chi}^{0}-\nu_{\tau}$. For 
a discussion of these decays and further references on the 
topic, we refer the reader to Ref.~\cite{CHUNneutr}.

\begin{figure}[ht] 
\begin{center} 
\epsfxsize= 7.92cm 
\leavevmode 
\epsfbox{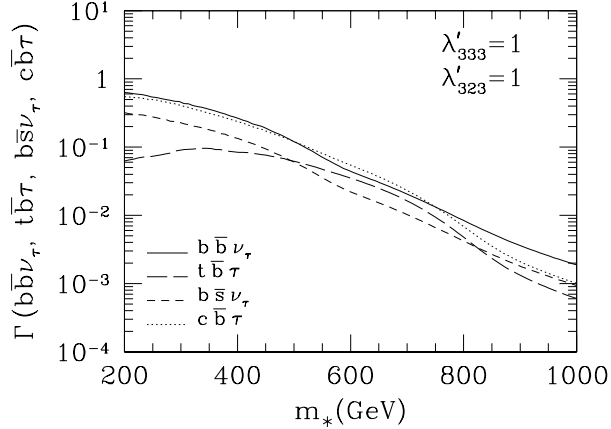}
\end{center} 
\vspace*{-0.5truecm}
\caption[f1]{\small{Widths (in GeV) for the four decays 
 $\tilde{\chi}^0_1\to b \bar{b} \nu_{\tau}$, $t \bar{b} {\tau^-}$,
 $b \bar{s} \nu_\tau$, and $c \bar{b}\nu_\tau$, when only the two 
 couplings $\lambda^\prime_{333}$ and $\lambda^\prime_{323}$ are
 nonvanishing, as function of the sfermion mass $m_\ast$, equal to
 $m_{\tilde{Q}}=m_{\tilde{U^c}}=m_{\tilde{D^c}}$, and to 
 $(1.4)\times m_{\tilde{L}}$ (with $m_{\tilde{L}}=m_{\tilde{E^c}}$), 
 for $\mu\!=\! 1500\,$GeV, $m_{\tilde{B}}=600\,$GeV, and
 $m_{\tilde{W}}=2 m_{\tilde{B}}$. The value of $\tan \beta$ is $3$ and
 all trilinear soft terms are chosen in such a way to have vanishing
 left-right mixing terms in the sfermion mass matrices
 squared. From Ref.~\cite{BGKT}.}}
\label{lplp} 
\end{figure}

The results may be summarized as follows, when the couplings
$\lambda^\prime_{333}$, or $\lambda^\prime_{332}$, and
$\lambda^\prime_{323}$ are considered.  Quite generically, if the
lightest neutralino is lighter (or only slightly heavier) than the
$t$-quark, it decays predominantly into massless fermions. Depending
on the relative size of the $R_p$-violating couplings, the dominant
decays are, {\it e.g}, $\tilde{\chi}^0_1 \to b \bar{b}\nu_\tau$, due to
a dominant $\lambda_{333}^{\prime}$ coupling, and 
$\tilde{\chi}^0_1 \to c \bar{b} \tau^-$, 
$\tilde{\chi}^0_1 \to s \bar{b} \nu_{\tau}$, and their CP-conjugate 
states, induced by $\lambda_{323}^{\prime}$ and
$\lambda_{332}^{\prime}$.  When two such couplings are simultaneously
present and of the same size, the rates for various massless states
are of the same order of magnitude over a wide range of parameter
values, see for example Fig.~\ref{lplp}. In this and the following
Figure, the lightest neutralino decay widths are plotted for values of
$\lambda^\prime_{333}$ and $\lambda^\prime_{323}$ equal to 1. Widths
for lower values of these couplings are obtained by scaling down those
shown in these Figures.

\begin{figure}[ht] 
\begin{center} 
\epsfxsize= 7.92cm 
\leavevmode 
\epsfbox{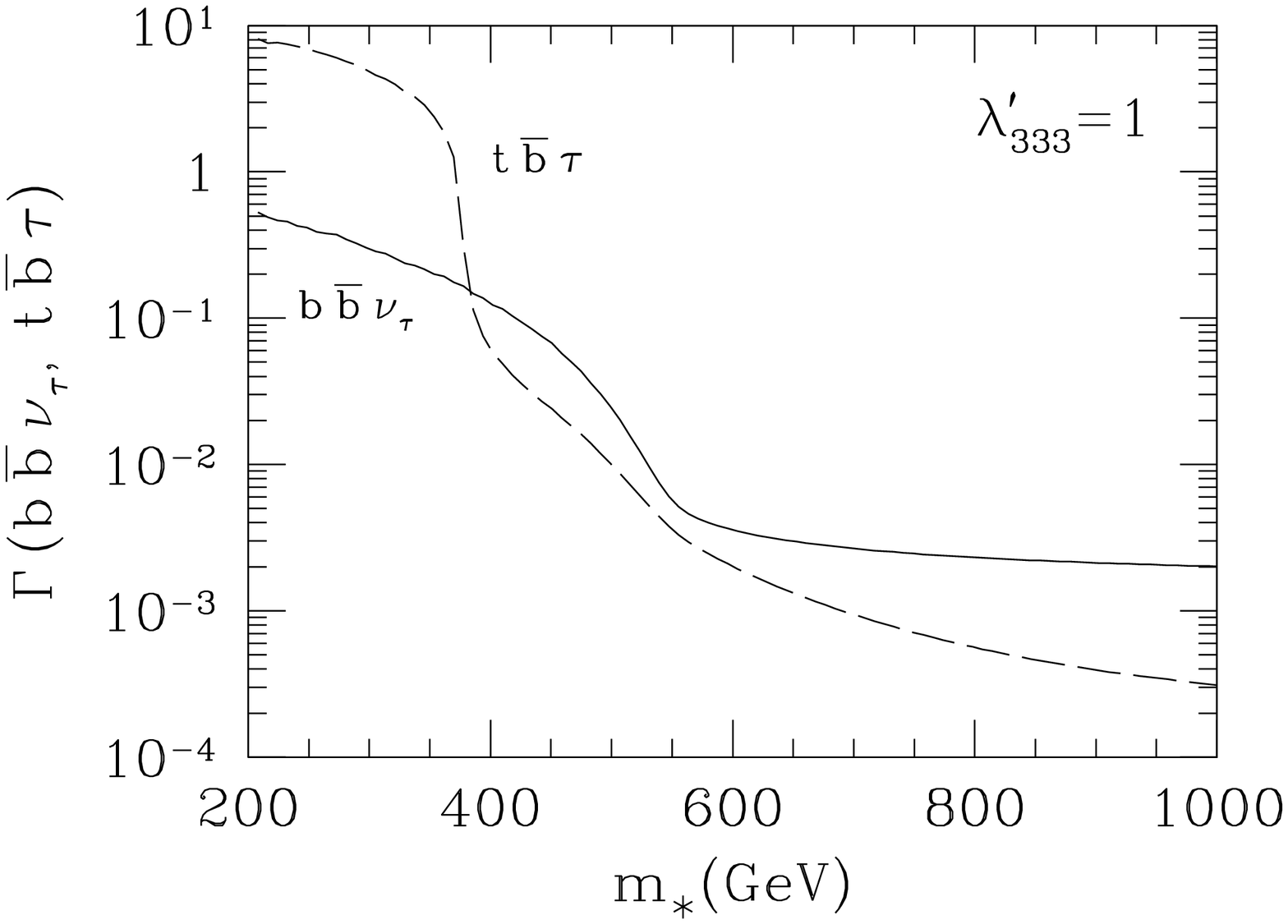} 
%\hspace*{0.4truecm}
\epsfxsize= 7.92cm 
\epsfbox{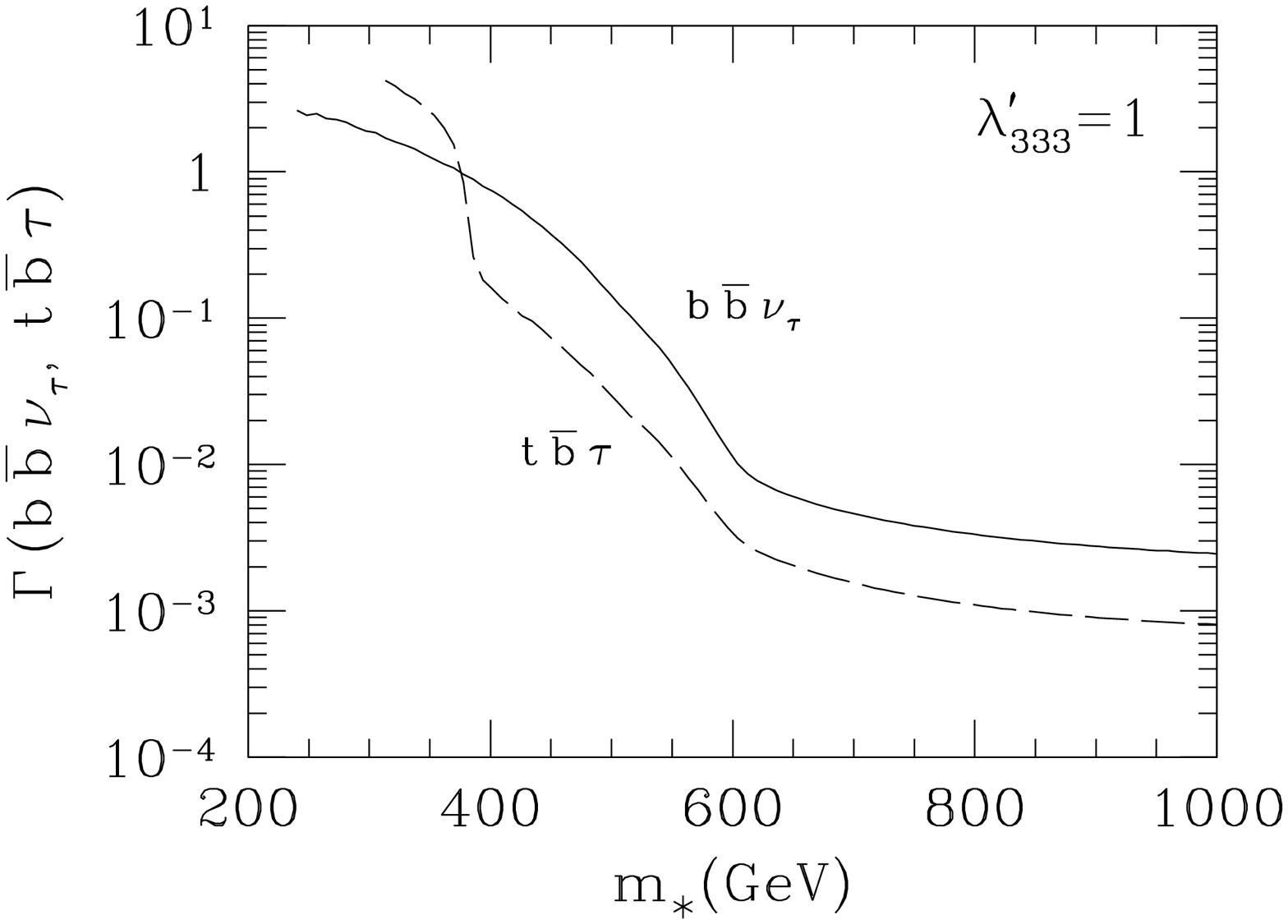} 
\end{center} 
\vspace*{-0.5truecm}
\caption[f1]{\small{$\Gamma(\tilde{\chi}^0_1\!\to\! b \bar{b} \nu_{\tau})$ 
 and $\Gamma(\tilde{\chi}^0_1\!\to \!t \bar{b} {\tau^-})$ (in GeV),
 respectively solid and long-dashed lines, versus
 $m_\ast\!=\!m_{\tilde{Q}}\!=\!m_{\tilde{U^c}}\!=\!m_{\tilde{D^c}}$, for
 $\mu\!=\! 600\,$GeV, $m_{\tilde{L}}\!=\!m_{\tilde{E^c}}\! =\!600\,$GeV,
 $m_{\tilde{B}}=600\,$GeV, and $m_{\tilde{W}}=2 m_{\tilde{B}}$. The
 value of $\tan \beta$ is $3$ in the left frame, $30$ in the right one. 
 All trilinear soft terms are fixed at $350\,$GeV in the left frame,
 and at $150\,$GeV in the right frame. From Ref.~\cite{BGKT}.}}
\label{binohino-mix} 
\end{figure}

When $\tilde\chi_1^0$ is substantially heavy, decays with the
$t$-quark in the final state, such as $t \bar{b}\tau^-$, 
$t \bar{s}\tau^-$, and their CP conjugated states, can become competitive. In
general, for moderate/large left-right mixing terms in the sfermions
virtually exchanged, and/or of substantial Bino-Higgsino mixing, these
decay modes can be large at not too large values of $\tan \beta$.
Although in general subdominant, their rate can be comparable to those
of massless modes, see Figs.~\ref{lplp} and~\ref{binohino-mix}.

Furthermore, an overall increase in the total decay width of
$\tilde{\chi}_1^0$ is observed, if $\tilde{\chi}_1^0$ is mainly a
Wino, as it may happen if gaugino mass unification is not imposed, or
in anomaly-mediated supersymmetry-breaking
scenarios~\cite{Gherghetta:1999sw}, or in grand-unification models
with an additional strong hypercolor
group~\cite{NEWUNIF}.

It should be noted here that, if $\tilde{\chi}_1^0$ is lighter than
the $t$-quark, the $\tilde{\chi}_1^0$ decays are not suitable to study
the lepton structure of the couplings $\lambda^\prime_{i33}$. In this
case, only decays with neutrinos in the final states are possible,
i.e.  $\tilde{\chi}_1^0 \to b \bar{b} \nu_i$. The modes in which the
charged leptons are present, $\tilde{\chi}_1^0 \to t \bar{b} \ell_i$
and $\tilde{\chi}_1^0 \to \bar{t} b \bar{\ell}_i$, are kinematically
forbidden.  A study of the lepton structure of these couplings could
reveal precious information on the origin of neutrino masses
(see discussion in Section~\ref{sleptnuphys}).
Unfortunately, independent collider probes are restricted to the
case of heavy neutralinos.  Couplings such as $\lambda^\prime_{i23}$,
which are far less relevant for neutrino physics, on the contrary, 
allow light final states, containing a charged
lepton, $\tilde{\chi}_1^0 \to c \bar{b} \ell_i$.

Chargino decays could be ideal to probe the $i$ dependence of the 
couplings $\lambda^\prime_{i33}$. The final states induced by 
such couplings, 
\begin{equation}
 \tilde{\chi}^-_1 \,\to\, b \bar{b} \ell_i  \,, \quad \quad 
 \tilde{\chi}^-_1 \,\to\, b \bar{t} \nu_i   \,, \quad \quad 
 \tilde{\chi}^-_1 \,\to\, t \bar{t} \ell_i  \,, 
\label{genchardecRP}
\end{equation}
where the last one is due to nonvanishing left-right mixing
terms in the sfermion mass matrix. Among these states, the first
one, in particular, could easily supply the information that
decays of a light neutralino cannot give. These $R_p$-violating 3-body
decays, however, must compete with the $R_p$-conserving 2-body decay
$\tilde{\chi}^-_1 \,\to\, W^- \tilde{\chi}^0_1$. In general, for 
$\tilde{\chi}^-_0$ lighter than $ \tilde{\chi}^-_1$, their rates are
much smaller than the rate of the dominant mode. For rather large
$R_p$-violating couplings, the $3$-body decays, although subleading,
may be nonnegligible. They may, however, be the dominant ones, when
the 2-body decay $ \tilde{\chi}^-_1 \,\to\, W^- \tilde{\chi}^0_1$ is
suppressed. This is typically the case of models in which
$\tilde{\chi}^-_1$ and $\tilde{\chi}^0_1$ are nearly degenerate. A
sufficient condition for this to happen is that $\tilde{\chi}^0_1$ is
mainly a Wino or mainly a Higgsino.
A detailed analysis of chargino decays in these 
scenarios will be presented elsewhere~\cite{CHARGINO}.

We conclude this Section by listing the chargino decays due to 
the coupling $\lambda^\prime_{i32}$ and $\lambda^\prime_{i23}$.
The coupling $\lambda^\prime_{i23}$ gives rise to the decays:
\begin{equation}
 \tilde{\chi}^-_1 \,\to\, b \bar{s} \ell_i  \,, \quad \quad 
 \tilde{\chi}^-_1 \,\to\, b \bar{c} \nu_i   \,, \quad \quad 
 \tilde{\chi}^-_1 \,\to\, t \bar{c} \ell_i  \,, \quad \quad 
 \tilde{\chi}^-_1 \,\to\, t \bar{s} \nu_i   \,,
\label{chardeci23}
\end{equation}
and all the CP-conjugated ones, 
whereas the decays obtained through the coupling 
$\lambda^\prime_{i32}$ are:
\begin{equation}
 \tilde{\chi}^-_1 \,\to\, b \bar{s} \ell_i  \,, \quad \quad 
 \tilde{\chi}^-_1 \,\to\, t \bar{s} \nu_i   \,,
\label{chardeci32}
\end{equation}
when the left-right mixing of the strange squark is assumed 
to be negligible.

\section{Collider- neutrino-physics interplay} 
\label{sleptnuphys}

The reasons for dwelling on $L$-violating couplings of type
$\lambda^\prime_{ijk}$ with third generation indices, is
that they have a particularly important role for neutrino spectra.

\begin{figure}[ht]
\begin{center}
\epsfxsize= 5.8cm
\leavevmode
\epsfbox{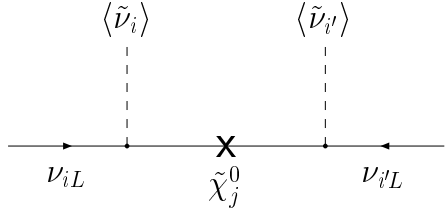}
\end{center}
\vspace*{-0.5truecm}
\caption[f1]{\small{The tree-level contribution to the Majorana 
 entry in the neutrino mass matrix, induced by sneutrino
 {\it v.e.v}'s. }}
\label{dnutreemass}
\end{figure}

It is well known that, additional $R_p$-violating interactions,
specifically interactions violating $L$, induce Majorana masses for
all neutrinos, without having to introduce new superfields in addition
to those present in a Minimal Supersymmetric Standard Model. In
generic $R_p$-violating models, neutrino may acquire mass at the tree
level and at the quantum level~\cite{generalNU}.  The tree-level
contributions are mainly governed by $R_p$-violating bilinear
couplings. The radiative contributions arise at the one- or the
two-loop level. At the one loop, they are controlled by trilinear
superpotential $R_p$-violating couplings and left-right sfermion
mixing mass terms. The two-loop contributions are due to trilinear
superpotential $R_p$-violating couplings, left-right sfermion mixing
terms, and to $R_p$-violating soft trilinear superpotential
terms~\cite{BL}.

The tree-level mass is originated by bilinear $R_p$-violating terms, 
in particular by the misalignment of superpotential bilinear terms, 
\begin{equation}
 W = \mu_i L_i H_U 
\quad\quad (i=1,2,3)
\,,
\end{equation}
and scalar potential bilinear terms, 
\begin{equation}
 V = B_i \tilde{L}_i H_U + m_{L_iH}^2 \tilde{L}_i H_U
\quad\quad (i=1,2,3)
\,.  
\end{equation}
(The symbol $H_U$ indicates the isospin $1/2$ doublet superfield 
as well as its scalar component.) This misalignment is expressed by
the nonvanishing value of the quantities:
\begin{equation}
  \Delta B_i       = B_i-B\,\hat{\mu}_i\,,   
\hspace*{0.5truecm}  \quad
  \Delta m_{L_iH}^2= m_{L_iH}^2-
      \left(m_{H_D}^2\!-\!m_{L_i}^2\right)\hat{\mu}_i\,,
\label{misalign}
\end{equation}
where $\hat{\mu}_i = {\mu_i}/{(\mu^2+\sum\mu_i^2)^{1/2}}$, 
and $\mu$ is the usual parameter in the superpotential term 
$\mu H_D H_U$. They induce vacuum expectation values
({\it v.e.v.}) for the sneutrinos, $v_i$. Typically,
\begin{equation}
 {m_{SUSY}}^2 \cdot v_i   \   \sim \ 
 \Delta B_i<\!H_U\!>+\Delta m_{L_iH}^2 <\!H_D\!> \,,
%  {m_{L_i}^2-\frac{1}{2}m_Z^2\cos 2\beta},
\label{vevs}
\end{equation}
where, $m_{SUSY}$ is, generically, the soft mass for the scalar 
component of the lepton doublet superfields plus $D$-term 
contributions.  In a
basis in which the terms $\mu_i L_i H_U$ in the superpotential are
rotated away, the tree-level contributions to the neutrino mass matrix
are given by:
\begin{equation}
 m_{\nu,ii^\prime} \sim 
    \frac{1}{2} \frac{g^2_j v_iv_{i^\prime}}{m_{\tilde{\chi}^0_j}}\,,
\label{neutrvevmass}
\end{equation}
and are obtained by the tree-level diagrams in Fig.~\ref{dnutreemass}.
Both $SU(2)$ and $U(1)$ gaugino mediate such a diagram.  A suppression
of these contributions is granted by rather small {\it v.e.v}'s and/or
cancellations among the two terms in Eq.~(\ref{neutrvevmass}) (i.e. 
$g_1^2/M_1 + g_2^2 /M_2=0$, with $M_1$ and $M_2$, the two gaugino
masses).

\begin{figure}[ht]
\begin{center}
\vspace*{0.5truecm}
\epsfxsize= 5.8cm
\leavevmode
\epsfbox{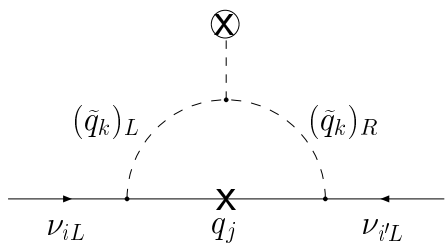}
\end{center}
\vspace*{-0.5truecm}
\caption[f1]{\small{A one-loop contribution to the Majorana
 neutrino mass arising from the $\Delta L = 1$ operators of the
 superpotential in Eq.~(\ref{superpot}).}}
\label{dnu1loopmass}
\end{figure}

The remaining neutrino mass terms are induced by trilinear 
$R_p$-violating terms through loops mediated by quarks-squarks and/or
leptons-sleptons.  A typical quark-squark loop
contribution is shown in Fig.~\ref{dnu1loopmass}.  Such a
contribution requires a chirality flip in the sfermion line. Choosing,
for simplicity, the basis in which left- and right- components of the
down quarks are diagonal, and assuming proportionality of the
left-right sfermion matrices to the corresponding Yukawa matrices, i.e. 
$(m^{2}_{\tilde{d}\,LR})_{ij} = {\cal M}^d_j (m_{d})_{j} \delta_{ij}$ 
and  
$(m^{2}_{\tilde{\ell}\,LR})_{ij}={\cal M}^l_j (m_{\ell})_{j}
\delta_{ij}$~\footnote{The effect of non-diagonal left-right mass 
 matrices on neutrino mass terms was considered in
 F.~Borzumati~{\it et al.} in Ref.~\cite{generalNU}.},  
the contributions to the neutrino mass matrix entries 
from quark-squark loops read:
\begin{equation}
 m_{\nu,ii^\prime} \sim \frac{3}{8\pi^2} \sum_{kj}  
   \lambda^\prime_{ikj}  \lambda^\prime_{i^\prime jk}
     (m_{d})_j (m_{d})_k \frac{{\cal M}^d_k}{m^2_{\tilde{d_k}}}, 
\label{neutrmassloop}
\end{equation}
where $m_{\tilde{d_k}}$ is a squark mass eigenvalue, and both
parameters ${\cal M}^d_k$ and $m_{\tilde{d_k}}$ are linked to the
mechanism of supersymmetry breaking. The analogous contributions from
lepton-slepton loops is obtained by replacing $\lambda^\prime_{ijk}$
with $\lambda_{ijk}$ in Eq.~(\ref{neutrmassloop}), and $m_d$ with
$m_l$.  The dominant among these contributions are those 
mediated by $b$-$\tilde{b}$ and by $\tau$-$\tilde{\tau}$.  Neutrino
mass terms $\ltap {\cal O} {\rm eV})$, imply that $\lambda_{i33}$ and
$\lambda^\prime_{i33}$ are smaller than $10^{-3}$-$10^{-4}$ for
supersymmetric masses of $100\,$GeV. If no special tuning among other
parameters occurs, {\it e.g.} if no hierarchy such as
${\cal M}^d_{3} \ll {\cal M}^d_{1} \simeq {\cal M}^d_{2}$ is realized, 
trilinear couplings such as
$\lambda^\prime_{i23}$ or $\lambda^\prime_{i32}$, and $\lambda_{i23}$
or $\lambda_{i32}$, 
induce loop contributions suppressed by $(m_{s}/m_{b})$ and
$(m_{\mu}/m_{\tau})$, with respect to the dominant ones.  The
suppression would be quadratic in the ratio of quark masses for the
one-loop contributions due to the couplings $\lambda^\prime_{i22}$.
Therefore, constraints obtained by imposing that the terms in
Eq.~(\ref{neutrmassloop}) are $\ltap {\cal O}(1 {\rm eV})$, are less
powerful for couplings involving one quark index of first or second
generation. It is interesting that, in general, collider physics
searches act exactly in the opposite way: direct constraints are
stronger for trilinear couplings with first and second generations
indices, than for those containing third generation indices.

The above formulas show how the predictability of generic
$R_p$-violating models is marred by the large number of parameters
contributing to the different neutrino mass matrix entries. To cope
with this difficulty, two different simplifying assumptions are in
general made. They give rise to two different classes of models, which
have been widely discussed in the literature.

The first is known as the class of bilinear $R_p$-violating models, in
which only bilinear terms exist~\cite{BILIN}.  By rotating away the
bilinear terms in the superpotential, the following trilinear
couplings are induced:
\begin{equation}
\lambda_{ijk}'= y^D_{jk} \,\hat{\mu_i}\,,
\hspace*{1truecm}
\lambda_{ijk} = y^L_{jk} \,\hat{\mu_i}
 \quad (i\ne j) \,,
\end{equation}
where $y^D_{jk}$ and $y^L_{jk}$ are, respectively, the down-quark and
charged-lepton Yukawa coupling matrices.  In general, the largest
neutrino mass term comes from the tree-level contribution. The second
largest mass terms are due to $b$-$\tilde{b}$ loops, which give rise to
entries in the neutrino mass matrix $m_{\nu,ii^\prime}$ proportional
to
$$
\sim 
(3/8\pi^2) \lambda^\prime_{i33} \lambda^\prime_{i^\prime 33} 
 m_b^2 {\cal M}^d_3/ m_{\tilde{b}}^2
\ = \ 
(3/8\pi^2) y_b^2 \,\hat{\mu}_i \,\hat{\mu}_{i^\prime}
 m_b^2 {\cal M}^d_3/ m_{\tilde{b}}^2\,.
$$ 
Other large mass terms are induced by $\tau$-$\tilde{\tau}$ loops:
$$
\sim 
(1/8\pi^2) \lambda_{i33} \lambda_{i^\prime 33} 
 m_\tau^2 {\cal M}^d_3/ m_{\tilde{\tau}}^2
\ = \ 
(1/8\pi^2) y_\tau^2 \,\hat{\mu}_i \,\hat{\mu}_{i^\prime}
 m_\tau^2 {\cal M}^l_3/ m_{\tilde{\tau}}^2\,, 
$$
with $i,i^\prime \ne 3$, due to the 
fact that the couplings $\lambda_{ijk}$ are 
antisymmetric in the first two indices.
The smallness of neutrino masses 
constrain the {\it v.e.v}'s to be small, and in particular the
$\mu_i$'s to be small 
\footnote{It is often 
 assumed that such a low-energy situation is obtained from 
 a model in which 
 $\Delta B_i$, $\Delta m_{L_iH}^2$ and all the trilinear 
 couplings are vanishing at some fundamental high scale.  
 Nonvanishing $\Delta B_i$ and $\Delta m_{L_iH}^2$ are induced
 by the parameters $\mu_i$ when evolving the model down to low scale, 
 whereas the trilinear terms remain vanishing. As explained above, 
 these result from a rotation of the superfields $L_i$ and $H_D$.  
 It is easy to show that in such a case the neutrino mass 
 matrix generated by the tree-level and the $b$-$\tilde{b}$-loop 
 contributions has 
 rank $1$, and that the second heaviest neutrino is generated 
 by $\tau$-$\tilde{\tau}$ loops.   
}.
As a consequence, the trilinear couplings are also rather small, and
so are the loop contributions to neutrino masses. 
The lightest neutralino decays mainly as 
$\tilde{\chi}^0_1 \to W^- \ell^+ $ induced by the mixing 
$\tilde{\chi}^0_1-\nu$ proportional to a neutral
slepton {\it v.e.v.}.  
No signal for single-production of charged sleptons is expected
in these cases.

Strictly speaking, however, since the {\it v.e.v}'s are an algebraic
sum of two terms, the individual contributions to the $v_i$'s do not
have to be small. If a certain amount of tuning of the individual
contributions is allowed, these {\it v.e.v}'s may be sufficiently
suppressed.  In principle the suppression may be even strong enough to
render the tree-level mass contribution of Eq.~(\ref{neutrvevmass})
irrelevant for neutrino physics. Depending on the amount of
suppression, the rates for the $3$-body decays of the lightest 
neutralino may be comparable to that for the $2$-body decay 
$\tilde{\chi}^0_1 \to W^- \ell^+$, or even larger.  In such a case,
the main source for neutrino mass terms are the loop contributions,
which must be duly suppressed.  Responsible for their suppression may
be parameters other than the trilinear couplings, i.e.  other than the
$\mu_i$'s. For example, nearly vanishing left-right mixing terms in
the sfermion mass matrices, may be as effective as very small
$\lambda$ and $\lambda^\prime$ parameters.  Moreover, it should be
noted that the dominant one-loop contributions to $m_{\nu,ii}$, i.e. 
those mediated by $b$-$\tilde{b}$ and $\tau$-$\tilde{\tau}$ are 
proportional to the combination of couplings
$\left(\lambda^\prime_{i33}\right)^2$ and 
$\left(\lambda_{i33}\right)^2$, respectively, and are therefore
sensitive to the phases of these couplings. Thus, it is possible
that partial cancellations among the $b$-$\tilde{b}$ loops and the
$\tau$-$\tilde{\tau}$ ones occur.  Further suppression of the
$b$-$\tilde{b}$ loops may come from large sbottom masses.  The
observation of single production of sleptons would strongly hint at
any of these possibilities, i.e. at unexpectedly large sbottom masses
(even for not too heavy sleptons), small left-right mixing terms in
the sfermion mass matrices, and/or special relations among the
couplings of type $\lambda$ and those of type $\lambda^\prime$.

The second class of models that has been considered in the literature
is that of models in which only trilinear $R_p$-violating terms exist
at some high scale~\cite{TRILIN}. In such models, bilinear terms are
induced by the trilinear couplings $\lambda$ and $\lambda^\prime$,
through the renormalization group equation. The misalignment term
$\Delta B_i$ and $\Delta m_{L_iH}^2$ and the sneutrino {\it v.e.v}'s
$v_i$ obtained at the electroweak scale are proportional to
$\lambda^\prime_{ijj}$ and $\lambda_{ijj}$, in a basis in which
down-quark Yukawa couplings are diagonal.  Barring accidental
cancellations among the different terms inducing the $v_i$'s, the
smallness of the tree-level contributions to neutrino masses constrain
these couplings to be small, if a nonnegligible mixing among left- and
right-mass terms for sfermions exists.  As a consequence, the leading
loop contributions are also suppressed and no significant yield for
the strahlung of charged sleptons in association with a $t$- (and
possibly a $b$-) quark is expected. As in the previous class of
models, this could be observed only if left-right sfermion mixing
terms are efficiently suppressed, and/or squark masses are heavy,
and/or phases in $R_p$-violation parameters induce cancellations among
the different loop contributions.  Other one-loop contributions to
neutrino masses are due to the trilinear couplings $\lambda_{ijk}$ as
well as $\lambda^\prime_{ijk} (i\ne j)$.  The constraints on these
last couplings are, in general, less limiting and one can expect to
observe single production of sleptons, as for example 
$pp\to t \tilde{\mu} X$ or $pp \to t\tilde{\mu} s X$, even without
having to advocate the vanishing of other parameters or the mutual
cancellation of different loop contributions.

\begin{figure}[ht]
\begin{center}
\vspace*{0.5truecm}
\epsfxsize= 6.7cm
\leavevmode
\epsfbox{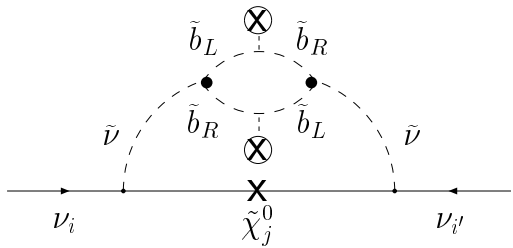}
\end{center}
\vspace*{-0.5truecm}
\caption[f1]{\small{A two-loop contribution to the Majorana
 neutrino mass arising from the $\Delta L = 1$ operators in the scalar
 potential in Eq.~(\ref{trilinearsoft}).}}
\label{dnu2loopmass}
\end{figure}

\begin{figure}[ht]
\begin{center}
\vspace*{0.5truecm}
\epsfxsize= 6.7cm
\leavevmode
\epsfbox{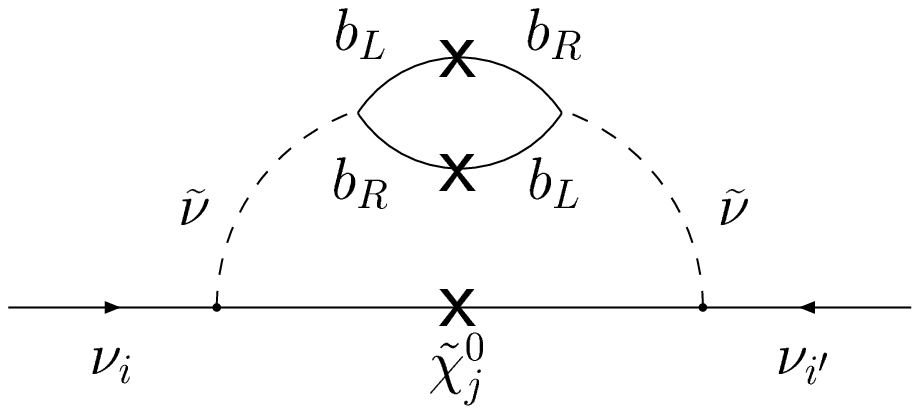}
\hspace*{1truecm}
\epsfxsize= 6.7cm
\leavevmode
\epsfbox{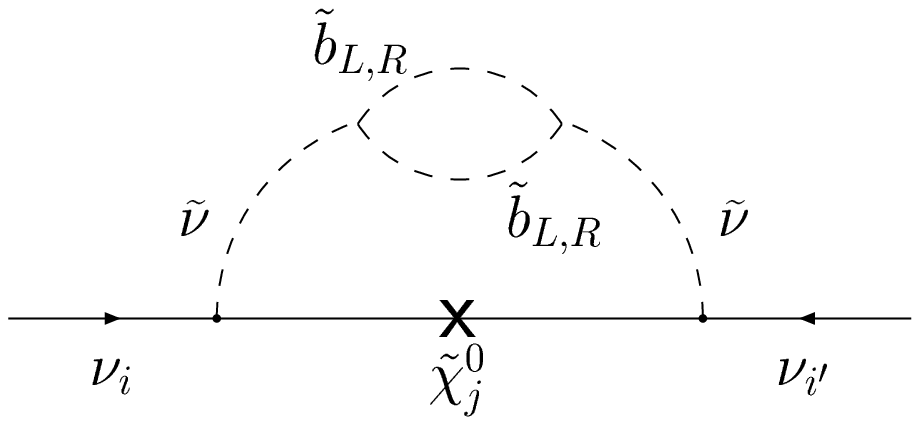}
\end{center}
\vspace*{-0.5truecm}
\caption[f1]{\small{Two-loop contributions to the Majorana
 neutrino mass involving the same trilinear superpotential couplings 
 $\lambda^\prime_{ijk}$ present in the one-loop diagram of 
 Fig.~\ref{dnu1loopmass}.}}
\label{dnu2loopmass-fer}
\end{figure}

Additional 2-loop contributions to neutrino masses also exists. They
are in general neglected, since assumed to be subleading with respect
to the one-loop contributions. The diagram with exchange of squarks
(and sleptons) in the upper loop, shown in Fig.~\ref{dnu2loopmass}, 
was already discussed in Ref.~\cite{BFPT}, in a framework in which
hard superpotential trilinear $R_p$-violating couplings are vanishing,
but soft trilinear $R_p$-violating couplings
\begin{equation}
 V =            A^\prime_{ijk} \tilde{L}_i \tilde{Q}_j \tilde{D^c}_k
 +  \frac{1}{2} A_{ijk} \tilde{L}_i \tilde{L}_j \tilde{E^c}_k
\label{trilinearsoft}
\end{equation} 
are generated by supersymmetry breaking.  (See also discussion in
Ref.~\cite{Frere:1999uv}.) There are also other diagrams with exchange
of quarks (and leptons) in the upper loop, shown on the left side of
Fig.~\ref{dnu2loopmass-fer}, and with exchange of squarks (and
sleptons) but no left-right mixing in the sfermion lines, as shown on
the right side of Fig.~\ref{dnu2loopmass-fer}.  If any of the
mechanisms advocated above to suppress the one-loop contribution to
neutrino mass (without suppressing the $\lambda$- and
$\lambda^\prime$-couplings) is present, these two-loop diagrams cannot
be neglected any longer~\cite{BL}.  

The sneutrino state $\tilde{\nu}$ has a CP-even and a CP-odd component
that are degenerate at the tree level. A splitting in the mass of
these two components is induced at the quantum level~\cite{sneusplitt}
by the upper loops in Figs.~\ref{dnu2loopmass}
and~\ref{dnu2loopmass-fer}. Because of this splitting, a nonvanishing
contribution to neutrino mass terms is originated.  The contribution
coming from the diagram of Fig.~\ref{dnu2loopmass} is:
\begin{equation}
 m_{\nu,ii^\prime} \sim \frac{3g_j^2}{(16\pi^2)^2} \,
    A^\prime_{i33}  A^\prime_{i^\prime 33}
    \frac{\left(m_{b}  {\cal M}_3^d\right)^2}{m^4_{\tilde{b}}}
    \frac{m_{\tilde{\chi}^0_j}}{m^2_{\tilde{\nu}}}\,.
\label{neutrtwoloopsc}
\end{equation}
The symbol ${\cal M}_3^d$ is that already used in
Eq.~(\ref{neutrmassloop}), i.e. ${\cal M}_3^d \,m_b $ is the
left-right mixing terms in the sbottom mass matrix.  Similarly, the
diagrams in Fig.~\ref{dnu2loopmass-fer} yields the contribution:
\begin{equation}
 m_{\nu,ii^\prime} \sim \frac{3g_j^2}{(16\pi^2)^2} \,
    \lambda^\prime_{i33}  \lambda^\prime_{i^\prime 33}
    m_{b}^2  \frac{m_{\tilde{\chi}^0_j}}{m^2_{\tilde{\nu}}}\,,
\label{neutrtwoloopfer}
\end{equation}
where we have dropped a logarithmic dependence on $m_{\tilde{b}}$.
For the overall coefficients in Eqs.~(\ref{neutrtwoloopsc}) 
and~(\ref{neutrtwoloopfer}), see Ref.~\cite{BL}.

One way of suppressing the one-loop contribution to neutrino mass
terms, and to evade constraints on the parameters $\lambda^\prime$, is
to assume that ${\cal M}^d_3 \simeq 0$. In such a case, obviously,
also the contribution from the diagram in Fig.~\ref{dnu2loopmass}
nearly vanishes. The contribution in Eq.~(\ref{neutrtwoloopfer}),
however, remain unsuppressed and yield therefore a constraint on the
couplings $\lambda^\prime_{i33}$~\cite{BL}:
\begin{equation}
  \lambda^\prime_{i33} \ltap 0.01 
   \left(\frac{m_{\tilde{\nu}}}{100\,{\rm GeV}}\right)
   \left(\frac{100\,{\rm GeV}}{m_{\tilde{\chi}^0}}\right)^{1/2}\,,  
\label{twolooplambdaconstr}
\end{equation}
where the $SU(2)$ gauge coupling was used for the numerical 
evaluation.  
This is quite robust since not affected by the size of squark 
masses. Similar constraints are obtained for the couplings 
$\lambda_{i33}$.

For such values of $\lambda^\prime_{i33}$, the possibility of
observing slepton strahlung at the LHC gets restricted to the
kinematical region $m_{\tilde{\ell}} \le m_t -m_b$, see the right
frame in Fig.~\ref{stausumcrsec}.  Wider reach of values for
$m_{\tilde{\ell}}$, i.e. beyond the $t$-quark mark, are possible if
sleptons are produced through the couplings $\lambda^\prime_{i32}$ or
$\lambda^\prime_{i31}$. Both, the one-loop and the two-loop
contributions to neutrino mass terms obtained in this case, are
suppressed with respect to those of Eq.~(\ref{neutrtwoloopfer}) by the
factors $(m_s/m_b)$ and $(m_d/m_b)$. The corresponding constraints on
these couplings get weakened by factors $(m_b/m_s)^{1/2}$ and
$(m_b/m_d)^{1/2}$. Similarly, also the couplings $\lambda_{ij3}$ and
$\lambda_{j3i}$ may remain unsuppressed.  As already observed in
Section~\ref{slstrahl}, the $2 \to 2$ process induced by
$\lambda^\prime_{i32}$ or $\lambda^\prime_{i31}$ give rise to sleptons
in association with a $t$-quark only.  The subsequent decay of the
sleptons through the couplings $\lambda_{ij3}$ or $\lambda_{j3i}$, may
give rise to an unsuppressed final state $t \tau^- \nu_j$
indistinguishable from the final state $t \tau^- \nu_\tau$ obtained
from a charged Higgs boson. Therefore, through $R_p$-violating
couplings, sleptons may contaminate the charged Higgs signal, even
when constraints from neutrino physics are imposed.

Last, but not least, it should be kept in mind that even the bounds in
Eq.~(\ref{twolooplambdaconstr}) and the corresponding one for the
coupling $\lambda_{i33}$ may be evaded. For this, it is sufficient
that special relations among the absolute value of
$\lambda^\prime_{i33}$ and $\lambda_{i33}$ and the value of their
phases induce a partial cancellation between the contribution in
Eq.~(\ref{neutrtwoloopfer}) against that from the similar $\tau$ and
$\tilde{\tau}$ diagrams. In this case, the possibility of observing 
slepton-strahlung induced by the coupling $\lambda^\prime_{i33}$ 
may remain unaffected by neutrino mass constraints.

In turn, it is possible that the one-loop contribution to neutrino
masses is, indeed, reduced by the smallness of the $\lambda^\prime$
and $\lambda$ couplings, while the left-right mixing terms in the
sfermion mass matrices remain sizable. In such a case, the neutrino
mass contribution from the two-loop fully scalar diagram in
Fig.~\ref{dnu2loopmass} induces strong constraints on the
$R_p$-violating soft couplings $A^\prime_{i33}$:
\begin{equation}
  A^{\prime}_{i33} {\cal M}^d_3  \ltap  100\,{\rm GeV}^2 
   \left(\frac{m_{\tilde{\nu}}}{100\,{\rm GeV}}\right)
   \left(\frac{100\,{\rm GeV}}{m_{\tilde{\chi}^0}}\right)^{1/2}
   \left(\frac{m_{\tilde{b}}}{100\,{\rm GeV}}\right)^2 \,,
\label{twoloopAconstr}
\end{equation}
and similar ones for the couplings $A_{i33}$.  Note that this is an
order of magnitude estimation. An explicit calculation, in which no
approximation is made on the loop function, may attenuate this
constraint.

\begin{figure}[ht]
\begin{center}
\vspace*{0.5truecm}
\epsfxsize= 4.3cm
\leavevmode
\epsfbox{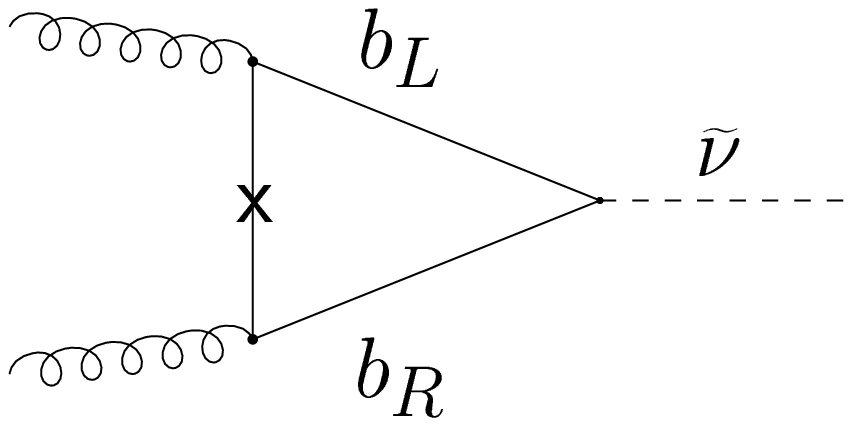}
\hspace*{0.5truecm}
\epsfxsize= 4.3cm
\leavevmode
\epsfbox{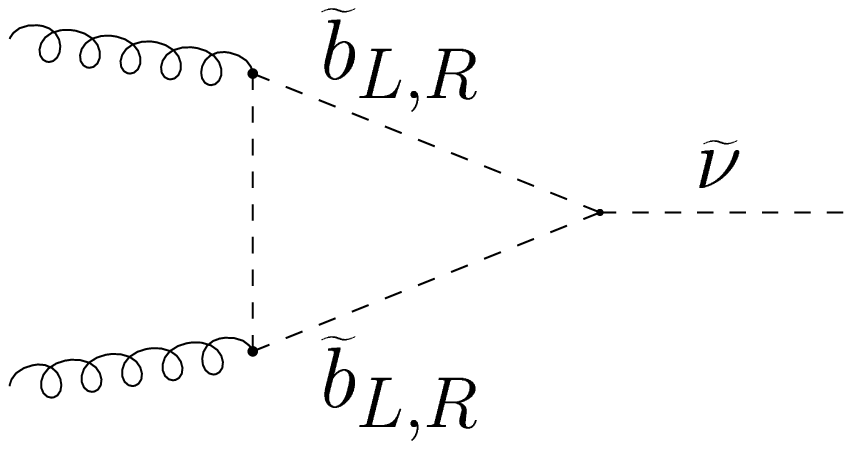}
\hspace*{0.5truecm}
\epsfxsize= 4.3cm
\leavevmode
\epsfbox{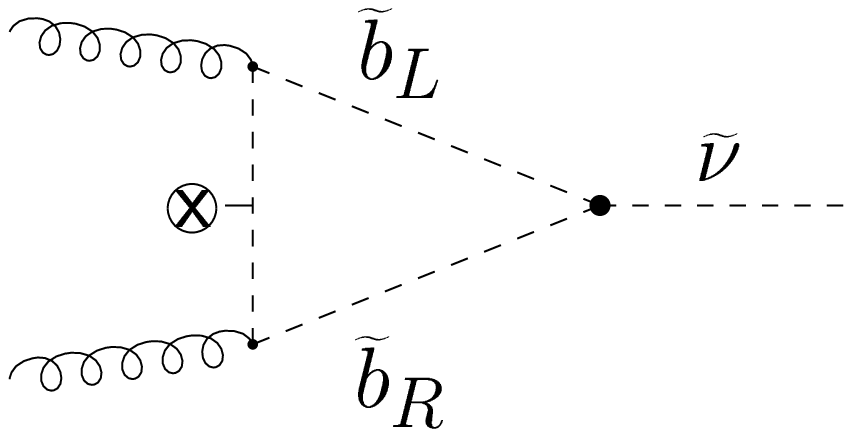}
\end{center}
\vspace*{-0.3truecm}
\caption[f1]{\small{Gluon-fusion production of sneutrinos.}}
\label{sneugluefusion}
\end{figure}

We conclude this Section by observing that constraints such as that in
Eq.~(\ref{twolooplambdaconstr}), will also affect the possible
production of sneutrinos through gluon
fusion~\cite{BKP,Bar-Shalom:1998xz,Chaichian:2001vi} and their
subsequent decays in two photons $\tilde{\nu} \to \gamma \gamma$.  It
has been suggested that sneutrinos in $R_p$-violating models may mimic
neutral Higgs bosons and spoil their observation and mass
determination.  They can be produced through the elementary process 
$b \bar{b} \to \tilde{\nu}$, as well as strahlung processes, or
through gluon fusion, as shown in Fig.~\ref{sneugluefusion}.  The
first two production mechanisms are induced by the couplings
$\lambda^\prime_{i33}$. These same couplings are also partially
responsible for the latter mechanism, as the first two diagrams in
Fig.~\ref{sneugluefusion} show.  The last diagram, on the contrary,
depends only on the couplings $A^\prime_{i33}$.

For these production mechanisms to be relevant, the generic
constraints from neutrino physics must be evaded.  Having assumed some
accidental cancellation of the tree-level neutrino contributions,
vanishing left-right mixing terms in the down-squark mass matrices are
enough to leave the couplings $\lambda^\prime_{i33}$ unconstrained by
the one-loop contribution to neutrino masses.  The bound from the
two-loop contribution given in in Eq.~(\ref{twolooplambdaconstr}),
however, may be enough to reduce the $b \bar{b} \to \tilde{\nu}$ and
strahlung production mechanisms to negligible levels. The same is true
also for the production through gluon fusion, since the first two
diagrams in Fig.~\ref{sneugluefusion} would be affected by small
couplings $\lambda^\prime_{i33}$ and the third by a vanishing
left-right mixing in the sbottom mass matrix.  In this case, a clean
detection of the neutral Higgs boson signal may be hoped for.  If on
the contrary, the one-loop contribution to neutrino mass are reduced
by tiny couplings $\lambda^\prime_{i33}$, and the left-right mixing in
the sbottom mass matrices are sizable, then single-production of
sneutrinos remains possible through the third gluon fusion diagram in
Fig.~\ref{sneugluefusion}. Parameters $A^\prime_{i33}$ of order 
$\sim 10\,$GeV may give smallish cross sections, but perhaps not so
small not to affect the neutral Higgs boson searches. Detailed
studies on this topic are lacking and are clearly called for.

It is obviously also possible that the one-loop contribution to
neutrino masses are subject to other type of suppressions, such as
the already mentioned cancellations among the $b-\tilde{b}$ and 
$\tau -\tilde{\tau}$ contributions. In such a case, the production 
of sneutrino at hadron collider may still be relevant. Once again, 
we can only conclude that direct and indirect searches should be 
pursed independently and should complement each other.

If neutrino physics constrains the $R_p$-violating trilinear couplings
to be below the $10^{-4}$-$10^{-3}$ mark, collider signals such as
charged slepton strahlung and sneutrino decays will be completely
negligible. Such small couplings would still be relevant for
phenomenological searches. They induce a decay of the lightest
neutralino, which would be a stable particle for vanishing values of
these couplings. It is in general assumed that the effect of such
couplings in other sectors would be completely irrelevant.  However,
in models with nearly-degenerate lightest chargino and lightest
neutralino, couplings as small as $10^{-5}$ may affect also the
phenomenology of the lightest chargino.  For such values of trilinear
couplings, the rate for the main decay mode of the lightest chargino
$\tilde{\chi}^-_1 \to W^- \tilde{\chi}^0_1 $ may become comparable or
even smaller (depending on the actual value of the mass difference
$m_{\tilde{\chi}^-_1} - m_{\tilde{\chi}^0_1} $ ) than that for the
$R_p$-violating chargino decay modes.  Nearly-degenerate
$\tilde{\chi}^-_1$ and $\tilde{\chi}^0_1 $ are present in scenarios in
which the lightest neutralino is mainly a Wino or a Higgsino. These
scenarios are realized in specific model of supersymmetry breaking,
such as anomaly-mediated supersymmetry-breaking
models~\cite{Gherghetta:1999sw}, or, for example, in grand-unification
models with an additional strong hypercolor group~\cite{NEWUNIF}.
For studies of the interesting signatures obtained in these 
cases, the reader is referred to Ref.~\cite{CHARGINO}.

\section{Conclusions and outlook}
\label{conclu} 
In spite of the impressive confirmation that the Standard Model has
obtained at LEP and at the Tevatron, the mechanism of electroweak
symmetry breaking is still elusive.  Searches for Higgs boson(s),
which are likely to mediate it, will play a pivotal role in collider
physics in the next coming years.  Some of the issues that these
searches will have to settle are if there exist only one neutral Higgs
boson or more, and if there are charged Higgs bosons, i.e. if there
are additional Higgs fields that are doublets of $SU(2)$. The
discovery of a charged Higgs boson will, undoubtedly, strengthen the
case for supersymmetry.  In order not to miss such an important
element of the Minimal Supersymmetric Standard Model at the LHC, it is
important that the cross section for its production is not too small,
so that even heavy masses may be reached.  As discussed in this
review, the charged Higgs strahlung is a suitable production mechanism
under this point of view: cross sections at the $fb$-level can be
obtained even for charged Higgs masses of $\sim 1\,$TeV.

The determination of the charged Higgs mass is also crucial to nail
down the parameters responsible for supersymmetry breaking.  The
cleanliness of this measurement, however, depends strongly on the fact
that there are no other particles mimicking the signals that charged
Higgs bosons produce.  In $R_p$-violating models, charged and neutral
sleptons may lead to an incorrect determination of the charged and
neutral Higgs boson masses.  The trilinear superpotential couplings
$\lambda^\prime_{i33}$ and $\lambda_{i33}$ can give LHC production
cross sections for charged and neutral sleptons very similar to that
of the corresponding charged and neutral Higgs bosons.

When the lepton index $i$ is of the third generation, they also give
rise to the same final states induced by the decays of Higgs
bosons. This is true, however, if these couplings are not much smaller
than numbers of ${\cal O}(1)$.  In the case of charged Higgs bosons
and sleptons radiated from a quark line, then, searches of final
states such as $(2t)\bar{b} (2\tau^-)X$, obtained for heavy
$\tilde{\chi}_1^0$ are necessary to establish whether the charged
Higgs signal was contaminated by charged sleptons through sizable
violation of the lepton number.  Moreover, when the leptonic indices
are of first or second generation, the presence of like-sign $e$'s and
$\mu$'s in possible final states such as $(2t)\bar{b} (2\mu^-)X$ or
$(2t)\bar{b} (2e^-)X$, will unambiguously indicate that charged
sleptons have been produced. In addition, also first and second
generation quark jets, induced for example by couplings
$\lambda^\prime_{ij3}$ or $\lambda^\prime_{i3j}$, are likely to arise
from charged sleptons and not from charged Higgs bosons.  Interesting
final states due to charged Higgs radiated from $t$-quark or $c$-quark
lines are, for example, $t \bar{t} s$, $(2t)\bar{s} (2\ell_i^-)X$, and
$(2c)\bar{b} (2\ell_i^-)X$, with $\ell_i^- = \tau^-, \mu^-, e^-$.

The neutral Higgs bosons at the LHC will be mainly produced through
gluon fusion, as well as strahlung off a top quark line.  Sneutrino
may also be produced in similar ways and with similar values of cross
sections thanks to ${\cal O}(1)$ couplings
$\lambda^\prime_{i33}$~\cite{Bar-Shalom:1998xz} as well as electroweak
scale massive couplings $A^\prime_{i33}$.  Also in this case, final
states overlapping with those of the neutral Higgs boson may be
obtained.

Contaminations between the Higgs bosons and sleptons signals become
negligible if $\lambda_{333}$, $\lambda^\prime_{333}$, and
$A^\prime_{i33}$ are sufficiently small.  In general, no strong upper
bounds have been obtained so far from direct searches of sparticle
production and/or particle/sparticle decays. Neutrino physics,
however, may play a crucial role in constraining such couplings. It
has been widely discussed in the literature how one-loop contributions
to neutrino masses impose that $\lambda^\prime_{i33}$ and
$\lambda_{i33}$ are not lareger than $10^{-4}$-$10^{-3}$. Since other
unknown parameters enter the one-loop calculation, however, it is
simple to evade these constraints.  Several possible ways have been
listed in this review. In addition, there exist two-loops
contributions to neutrino masses that are in general left out of
discussion, since considered subleading. Nevertheless, the
combinations of various parameters entering in the calculation of the
two-loop diagrams are different from those encountered in the calculation
of the one-loop diagrams. Therefore, the two-loop diagrams may not be
affected by the one-loop constraints. Although suffering a stronger
loop-suppression with respect to the one-loop diagrams, they may still
give too large contributions to neutrino masses. As it is explicitly
shown in this review, they yield constraints on the couplings
$\lambda^\prime_{i33}$ and $\lambda_{i33}$ that are more difficult to
evade (although not impossible), once specific choices of parameters
are made to cancel the one-loop contributions.  It turns out that
$\lambda^\prime_{i33}$ and $\lambda_{i33}$ are constrained to be a few
$\%$, whereas the couplings $A^\prime_{i33}$ should not exceed the
$10\,$GeV order of magnitude.

Values of $\lambda^\prime_{i33}$ at the percent level, can still allow
nonnegligible yields of charged sleptons at the LHC, although the
production cross section is considerably smaller than that for a
charged Higgs boson of equal mass.  Their detection, challenging,
relies on the search for final states different than those produced by
charged Higgs bosons. Sneutrinos may still be produced at the LHC,
through gluon fusion and/or strahlung off a quark line.  Detailed
studies of the impact of the parameters $A^\prime_{i33}$ and of the
distinguishability of sneutrinos from Higgs bosons are still missing.

Lepton violating couplings $\lambda^\prime_{i3j}$,
$\lambda^\prime_{ij3}$, with the quark index $j$ of second and
especially first generation, are much less affected by neutrino 
physics. The same is true for the couplings $\lambda_{ij3}$ and 
$\lambda_{j3i}$.  As it was extensively discussed in
Sections~\ref{chiggs} and~\ref{slstrahl}, it should be noted that the
$2 \to 2 $ elementary processes 
$g \bar{s}, g \bar{d} \to t \tilde{\ell}_i$ induced by the couplings
$\lambda^\prime_{i3j}$, do not need to be combined with the $2 \to 3$
processes $q\bar{q}, gg \to t \bar{s} \tilde{\ell}_i$ or 
$q\bar{q}, gg \to t \bar{d} \tilde{\ell}_i$ if the produced sleptons
$\tilde{\ell}_i$ decay leptonically. This is exactly the same
situation encountered in the case of strahlung-produced charged Higgs
bosons that decay into $\tau^- \nu_\tau$. Moreover the $2 \to 2$ cross
section for slepton-strahlung due to the quark-flavour violating
couplings $\lambda^\prime_{i3j}$, is bound to be larger than that
obtained from a coupling $\lambda^\prime_{i33}$ of equal strengh, due
to the larger content of the $d$- and the $s$-quark in the proton.
Therefore the subsequent decay of such sleptons into $\tau^- \nu_j$
induced by the couplings $\lambda_{ij3}$ and $\lambda_{j3i}$ will
provide a final state identical to that obtained from the charged
Higgs.  Sleptons of mass very close to that of the charged Higgs,
would therefore affect the search of charged Higgs. A considerable
splitting among the mass of these particles may lead to exciting
results at the LHC.

\vspace*{0.5truecm}
\noindent 
{\bf Acknowledgements} \\ 
Discussions with K.~Hagiwara and W.~Porod, as well as collaborations
with R.~Godbole, J.L.~Kneur, and N.~Polonsky are acknowledged.  F.~B.
thanks KEK, in particular K.~Hagiwara and S.~Choi for partial
financial support for this conference.  F.~T. acknowledges the
hospitality of the Theory Groups at KEK and at the Physics Department
of the University of Tokyo, where parts of the work reported here were
carried out.  The work of J.S.~L. was supported by the Japanese
Society for Promotion of Science, that of F.~B. was partially supported
by a Japanese Monbusho Fellowship and partially by the Japanese
Society for Promotion of Science.  Last, but not least, Y.~Fujii,
K.~Hagiwara and R.~Yahata are gratefully thanked for the impeccable
organization of the conference, and for taking care of many ``last
moment'' problems.

%\newpage

\end{document}